\documentclass[12pt,psfrag,tightenlines,eqsecnum,floats,aps,amsmath,amssymb,nofootinbib,prd,shownopacs,floatfix]{revtex4}

\usepackage{setspace}
\usepackage{amssymb}
\usepackage{amsmath,amssymb,amsfonts,amsthm,mathrsfs}
\usepackage[latin1]{inputenc}
\usepackage{graphicx,wrapfig}
\usepackage{enumerate} 
\usepackage{color}
\usepackage{mathrsfs}
\usepackage{subfigure}
\definecolor{mg}{rgb}{0.0, 0.5, 0.0}

\def\be{\nopagebreak[3]\begin{equation}}
\def\ee{\end{equation}}
\def\ba{\nopagebreak[3]\begin{eqnarray}}
\def\ea{\end{eqnarray}}
\newcommand{\f}{\frac}
\def\rmd{\rm d}
\def\lp{\ell_{\rm Pl}}
\def\mp{m_{\rm Pl}}

\def\t{\tilde}
\def\h{\hat}

\def\Ssc{\mathcal{S}_{\rm sc}}
\def\r{\mathfrak{r}_{\rm{curv}}}
\def\Kr{\mathcal{K}}
\def\scri{\mathcal{I}}
\def\scrim{\mathcal{I}^{-}}
\def\scrip{\mathcal{I}^{+}}
\def\DH{\emph{DH}\, }
\def\DHs{\emph{DHs}\,\,}
\def\TDH {\emph{T-DH}\, }
\def\ATDH {\emph{AT-DH}\, }

\begin{document}

\title{Black Hole evaporation: \\A Perspective from Loop Quantum Gravity}

\author{Abhay Ashtekar}
\affiliation { Institute for Gravitation and the Cosmos, Penn State University,
University Park, PA 16801.}
\begin{abstract}
A personal perspective on the black hole evaporation process is presented using as guidelines  inputs from:  (i) loop quantum gravity, (ii) simplified models where concrete results have been obtained, and, (iii) semi-classical quantum general relativity. On the one hand, the final picture is conservative in that there are concrete results that support each stage of the argument, and there are no large departures from general relativity or semi-classical gravity in tame regions outside macroscopic black holes. On the other hand it argues against certain views that are commonly held in many quarters, such as: persistence of a piece of singularity that constitutes a part of the final boundary of space-time; presence of an \emph{event} horizon serving as an absolute barrier between the interior and the exterior; and  the (often implicit) requirement that purification must be completed by the time the `last rays' representing the extension of this event horizon reach $\mathcal{I}^{+}$. 

\end{abstract}
\maketitle

\section{Introduction}
\label{s1}

Fig. 1(a) is the standard Penrose diagram of the gravitational collapse of a star to form a non-rotating black hole in classical general relativity (GR). While $\scrim$ serves as the past boundary of this space-time, $\scrip$ constitutes only a part of the future boundary which \emph{also includes the future, space-like singularity}. Therefore, while data for a zero rest mass field on $\scrim$ suffices to determine the field everywhere in space-time in the forward (in time)  evolution, data on $\scrip$ is clearly insufficient in the backward evolution; the singularity `soaks up part of the information' in generic solutions to field equations.  

\begin{figure} 
  \begin{center}
    \begin{minipage}{3in}
      \begin{center}
        \includegraphics[width=1.6in,height=2.5in,angle=0]{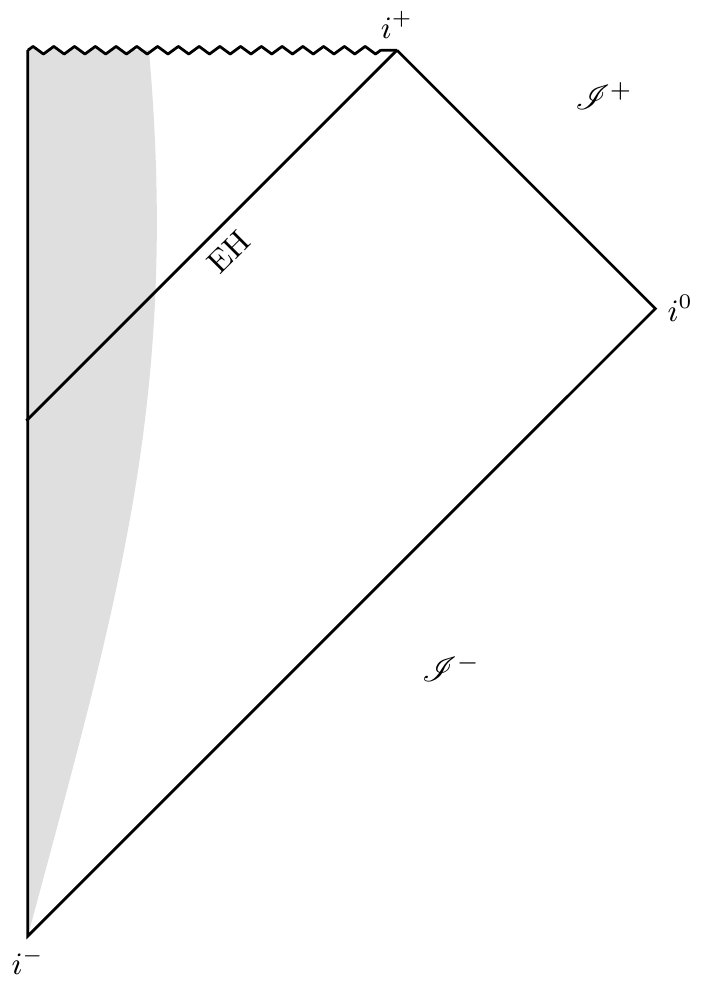} \\ {(a)}
         \end{center}
    \end{minipage}
    \begin{minipage}{3in}
      \begin{center}
       \includegraphics[width=1.7in,height=2.9in,angle=0]{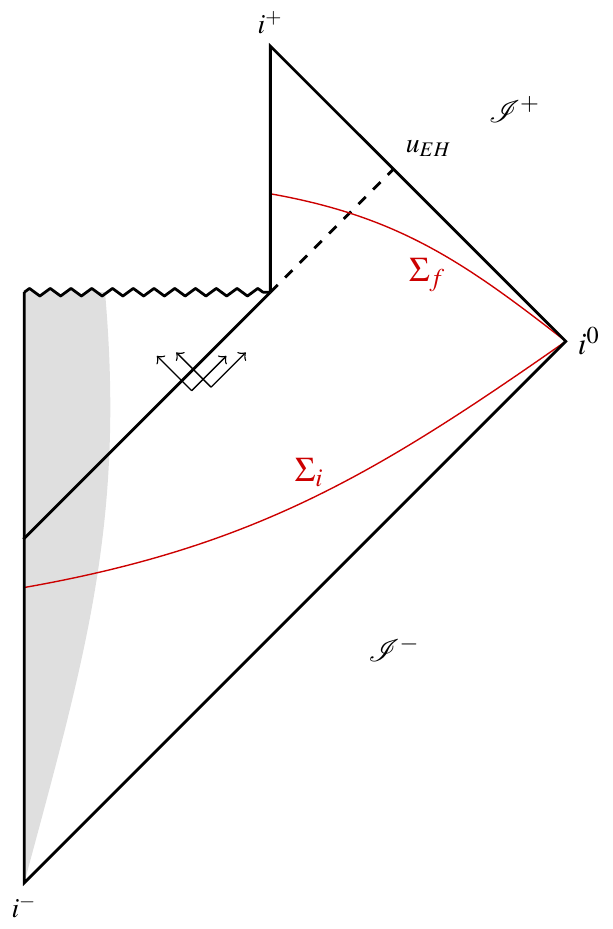} \\ {(b)}
        \end{center}
    \end{minipage}
\caption{\footnotesize{\emph (a): Gravitational collapse of a star. While the past boundary of space-time consists only of $\scrim$, the future boundary is the union of $\scrip$ {\it and the black hole singularity}. There is information loss in the future evolution because the singularity can soak up part of the information.  \\ (b) Commonly used Penrose diagram to depict black hole evaporation, including back reaction. Modes are created in pairs, one escaping to $\scrip$ and its partner falling into the black hole. The dashed line is the continuation of the Event Horizon that meets $\scrip$ at retarded time $u_{EH}$. If this were an accurate depiction, the future singularity would again act a sink of information.}} 
\label{fig:1}
\end{center}
\end{figure}

Hawking's celebrated 1975 calculation \cite{swh}  was carried out on this space-time, ignoring the back reaction. However, he also drew the Penrose diagram he expected to result once the back reaction is appropriately taken into account, shown in Fig. 1 (b). This diagram is routinely used in many discussions of the black hole evaporation process and the associated issue of potential information loss.  It depicts the idea that, as the back hole loses energy through radiation to $\scrip$, the `strength' of the singularity --captured by, say,  the Kretschmann scalar that goes as $(GM/r^{3})^{2}$ in spherically symmetric space-times--  diminishes and so the singularity ends, rather than extending all the way to $\scrip$ as in Fig. 1 (a).  Fluctuations in the vacuum at $\scrim$ are excited resulting in the creation of pairs of modes, one escaping to $\scrip$ and its partner falling into the singularity. If this were an accurate depiction of the process, then evolution from an early time (or initial) Cauchy surface $\Sigma_{i}$ to a late time surface (or final) $\Sigma_{f}$  would not be unitary because the partner modes disappear into the black hole. Therefore, Hawking was led to introduce a generalization of the standard (unitary) S-matrix to a \$-operator that can evolve pure states to mixed \cite{swh2}.  By and large, there is consensus in the relativity community that this conclusion would be inescapable if Fig. 1 (b) were to  faithfully capture the full evaporation process. However, it is important to note that Hawking's original space-time diagram \cite{swh} was not based on a detailed calculation; rather, it expressed his then expectation on how evaporation would proceed.  Interestingly, by 2016 his expectation changed completely. A new Penrose diagram was proposed  in \cite {hps} (their Fig. 2) in which there is no final singularity. In such a space-time the obvious obstruction to unitarity disappears.

However, Fig. 1(b)  is deeply instilled in much of the community and discussion of the issue of information loss tends to be based on the ensuing intuition. Indeed, the figure constitutes a main pillar in arguments  made by \emph{both} camps: those  in favor of information loss  in the asymptotic region of Fig. 1(b) (see e.g. \cite{uw}),  and those in favor of unitarity  in this space-time (see, e.g., \cite{apms,giddings,marolf}).  In the first case one is often led to baby universes. Explorations in the second camp tend to assume that unitarity would be restored because  `the full information will come out'  by the retarded time $u=u_{\rm EH}$, at which the null rays that extend the event horizon intersect $\scrip$ as in Fig. 1(b). Not surprisingly, one is then led to uncomfortable conclusions such as violation of semi-classical gravity in tame regions in which space-time curvature is far from that at the Planck scale. In particular, arguments that led to the introduction of  quantum xerox machines,  firewalls, and potential mechanisms based on fast scramblers are motivated by the paradigm in which the `purification occurs before $u_{\rm EH}$'  --although this assumption is often implicit because this premise is taken to be obvious.
 
 Loop quantum gravity suggests a different space-time diagram  that can lead to a unitary quantum evolution, where nothing dramatic happens in the low curvature, tame space-time regions. This is because several assumptions that led to Fig. 1(b) are violated. This new space-time diagram  and its implications will be discussed in Section \ref{s2}. Its salient features are: \vskip0.15cm
 \begin{figure} 
  \begin{center}
    \begin{minipage}{2.5in}
      \begin{center}
        \includegraphics[width=2in,height=3.4in,angle=0]{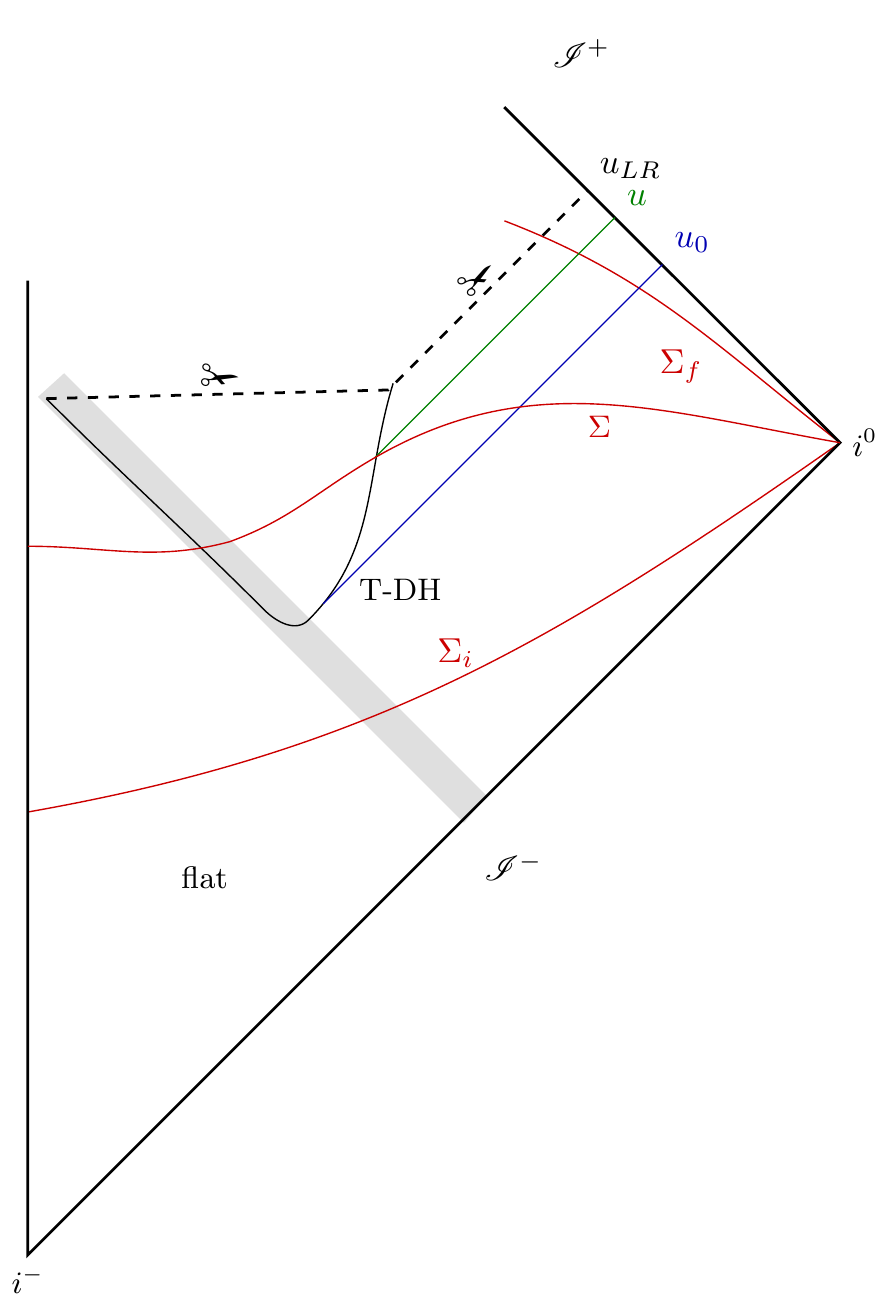} \\ {(a)}
         \end{center}
    \end{minipage}
   \hspace{.5in}
    \begin{minipage}{3in}
      \begin{center}
       \includegraphics[width=1.9in,height=3.7in,angle=0]{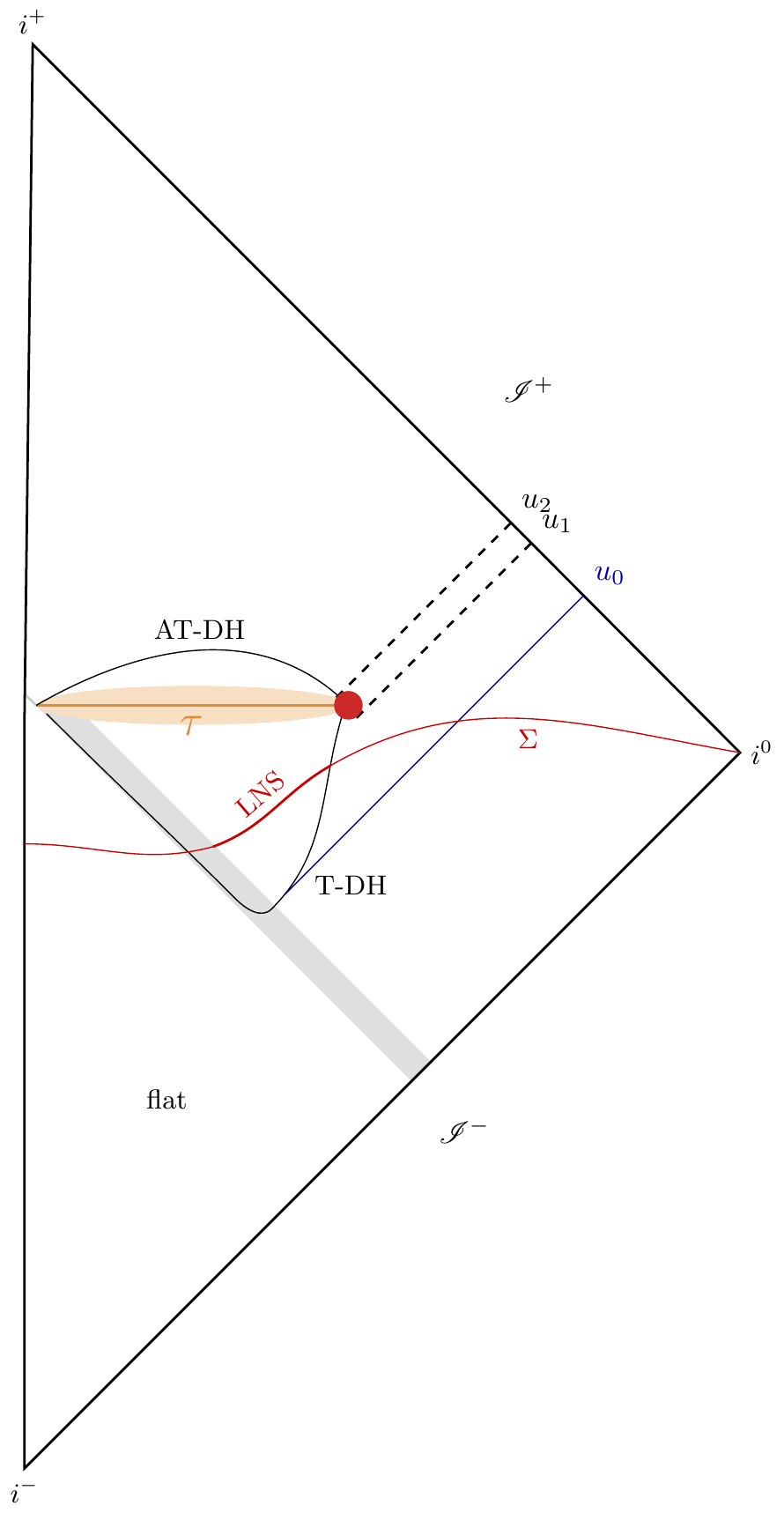} \\ {(b)}
        \end{center}
    \end{minipage}
\caption{\footnotesize{ (a) Part of space-time in which the semi-classical approximation holds.  The shaded region depicts an infalling massless scalar field that would give rise to a singularity in classical GR. But LQG quantum geometry effects intervene to prevent the formation of this singularity. This quantum geometry region and its future are excised (dashed line with scissors).  A trapping Dynamical horizon \TDH develops which is space-like with increasing area (in the outward direction) during the collapse, and time-like with decreasing area (in the future direction) during evaporation. Null rays from the contracting part of the \TDH go out to meet $\scrip$. Hawking radiation starts in earnest at $u=u_{0}$. The dashed line with scissors that includes the last ray $u =u_{LR}$ represents the future boundary of the semi-classical region.  Evolution from the initial Cauchy surface $\Sigma_{i}$ to generic Cauchy surfaces $\Sigma$ that lie entirely in the semi-classical region is unitary.  However evolution from $\Sigma$ to the portion of $\Sigma_{f}$ that lies within the semi-classical region is not.\\
 (b)  Proposed quantum extension of the space-time.  The classical singularity is replaced by a \emph{Transition surface} $\tau$, to the past of which we have a trapped region, bounded in the past by a trapping dynamical horizon \emph{T-DH}, and to the future of which we have an anti-trapped region bounded by an anti-trapping dynamical horizon \emph{AT-DH}. Cauchy surfaces $\Sigma$ develop astronomically long necks already in the semi-classical region and the long necks persist also in the quantum region. Partner modes that fall in the trapped region get stretched enormously and are slowly released to the future of the quantum region. Thus, `most of the purification' occurs to the future of the transition surface $\tau$. The dark (red) blob at the right end of $\tau$ is a genuinely quantum region discussed in sections \ref{s2.3} and \ref{s2.4}.}}
\label{fig:2}
\end{center}
\end{figure}

(1)  \emph{Dynamical Horizons}: What forms and evaporates is a dynamical horizon \cite{ak1,ak2,bbgvdb,akrev,boothrev}. There is no null event horizon causally isolating the `interior' region as in Fig. 1(a). During formation of the black hole, the trapping dynamical horizon \TDH  is space-like  and the marginally trapped surfaces (MTSs) on it grow in area in the outward direction, while during Hawking evaporation, \TDH is time-like and  the MTSs  shrink in area in the future direction.  As we will discuss in Section \ref{s2}, the absence of an event horizon and growth and shrinking of the \TDH are realized in detail in the case of a solvable spherical black hole model.  \vskip0.15cm
 
\noindent (2) \emph{ Non-trivial geometry in the trapped region:} Already in the semi-classical region, where curvature is much smaller than the Planck scale, the interior space-time geometry
{develops astonishing features}  \emph{once the back reaction due to the (negative energy)  Hawking flux is included.} For a macroscopic black hole, the interior geometry evolves very slowly but over an \emph{extremely} long time. If one introduces a geometrically natural foliation of  the interior --say slices with constant values of the Kretschmann  scalar, or of the area radius $r$ of round 2-spheres \cite{ao,aa-ilqg} --  one finds that these slices develop \emph{astronomically long necks}.  (These are the \emph{long neck surfaces}  depicted by the portion LNS of the Cauchy surface $\Sigma$ of Fig. 2(b).)  Now,  partners of the modes that escape to $\scrip$ enter the interior of the \TDH and evolve on this time-dependent background. One expects that, as in cosmology of the early universe, the astronomical expansion of necks would stretch their wave lengths enormously whence, when sufficient time has passed,  the interior of the \TDH can house an enormous and increasing number of these modes but with very small  total energy.  {A priori,} there is no obstruction to continued entanglement between the modes outside and those inside \emph{well beyond}  the Page time \cite{page}.
 \vskip0.15cm
 \noindent (3) \emph{Singularity resolution:} In LQG, the Big-Bang  singularity has been shown to be resolved in a robust manner in a large number of models (see, e.g., \cite{asrev,iapsrev}). For the Schwarzschild-interior, there is now an effective description of quantum geometry with several physically attractive features  that are hallmarks of the LQG quantum geometry \cite{aoslett,aos}. In particular,  we again have singularity resolution;  there are universal upper bounds on curvature invariants;  and the quantum modifications of Einstein's equations die very quickly away from the Planck regime. In the quantum extended space-time,  the singularity is replaced by a quantum `transition surface'  $\tau$ to the past of which we have a trapped (BH-type) region and to the future of which there is a (white-hole type) anti-trapped region 
 \footnote{I would like to emphasize that there is neither a black hole nor  a white hole, because both of them have singularities, while the quantum space-time has none. Therefore we will refer to these regions simply as {\it trapped} and {\it anti-rapped.} The past boundary  of the trapped region is a Dynamical Horizon  and its future boundary is the Transition Surface $\tau$. The past boundary  of the anti-trapped region is  $\tau$ and the future boundary is another Dynamical Horizon. See Fig. 2(b).}
(see Fig.  2(b)).  In this scenario, the quantum extension of space-time has an asymptotic region with $\scrip$.  The extremely long wave-length modes inside the \emph{T-DHs}\,  --that are entangled with those those emitted to $\scrip$ --  \,travel across the transition surface $\tau$ to the anti-trapped region and then leak out slowly  to $\scrip$ on a time scale  $\gtrsim M^{4}$ \cite{ori4}. Thus,  the correlations between outgoing modes (with an `approximately thermal' character) that were emitted in the `Hawking phase' of quantum radiation and their parters can be restored on $\scrip$ but on a much longer timescale. 
\vskip0.2cm

Each of these three ideas constitutes a major conceptual shift with respect to Fig. 1(b). Together they provide a more concrete basis for the general, LQG-based paradigm of  the black hole evaporation process  first proposed in \cite{aamb}, and discussed in a more concrete form in \cite{aa-ilqg}. Section \ref{s2} presents the conceptual underpinning of these ideas in greater detail and also discusses further issues that were not included in the summary sketched above. That discussion should make the current status of the program clearer. Similarities and differences with some of the other scenarios \cite{apms,ori1,uw,marolf,eh} are illustrated in Section \ref{s3}. In particular, there is a discussion of issues related to space-time diagrams compatible with geometrical considerations, proposed over the last decade or so (see in particular,  \cite{hayward,frolov,bardeen,ebms,crfv,ebtdlms,hhcr,bcdhr,mdcr}).

\section{Black hole evaporation: Inclusion of back-reaction}
\label{s2}

\begin{figure} 
  \begin{center}
    \begin{minipage}{2.5in}
      \begin{center}
        \includegraphics[width=1.5in,height=2.2in]{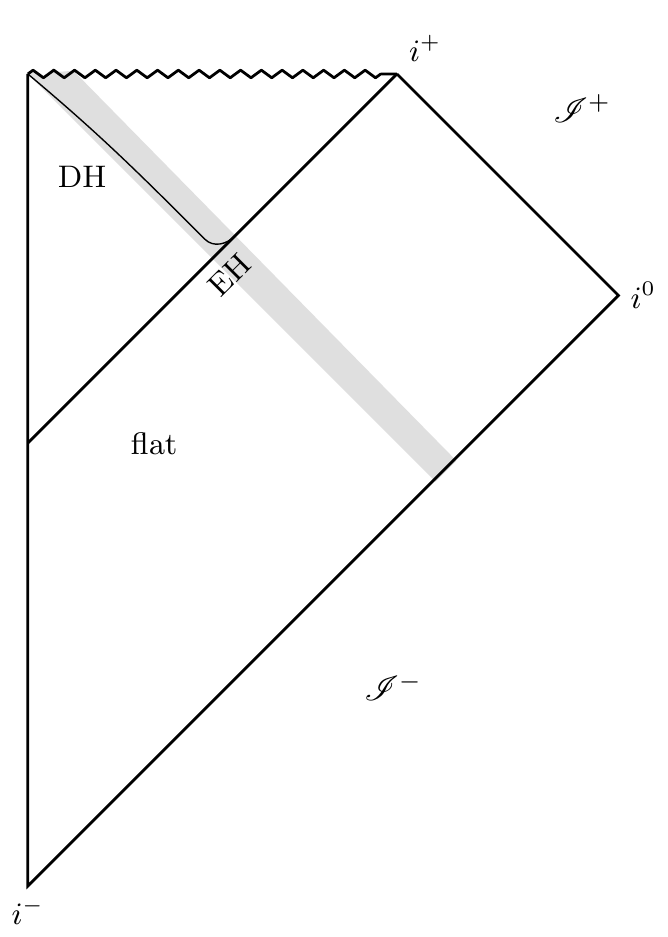}\\ {(a)}
        \end{center}
    \end{minipage}
   \hspace{.5in}
    \begin{minipage}{3in}
      \begin{center}
       \includegraphics[width=2.2in,height=2.5in]{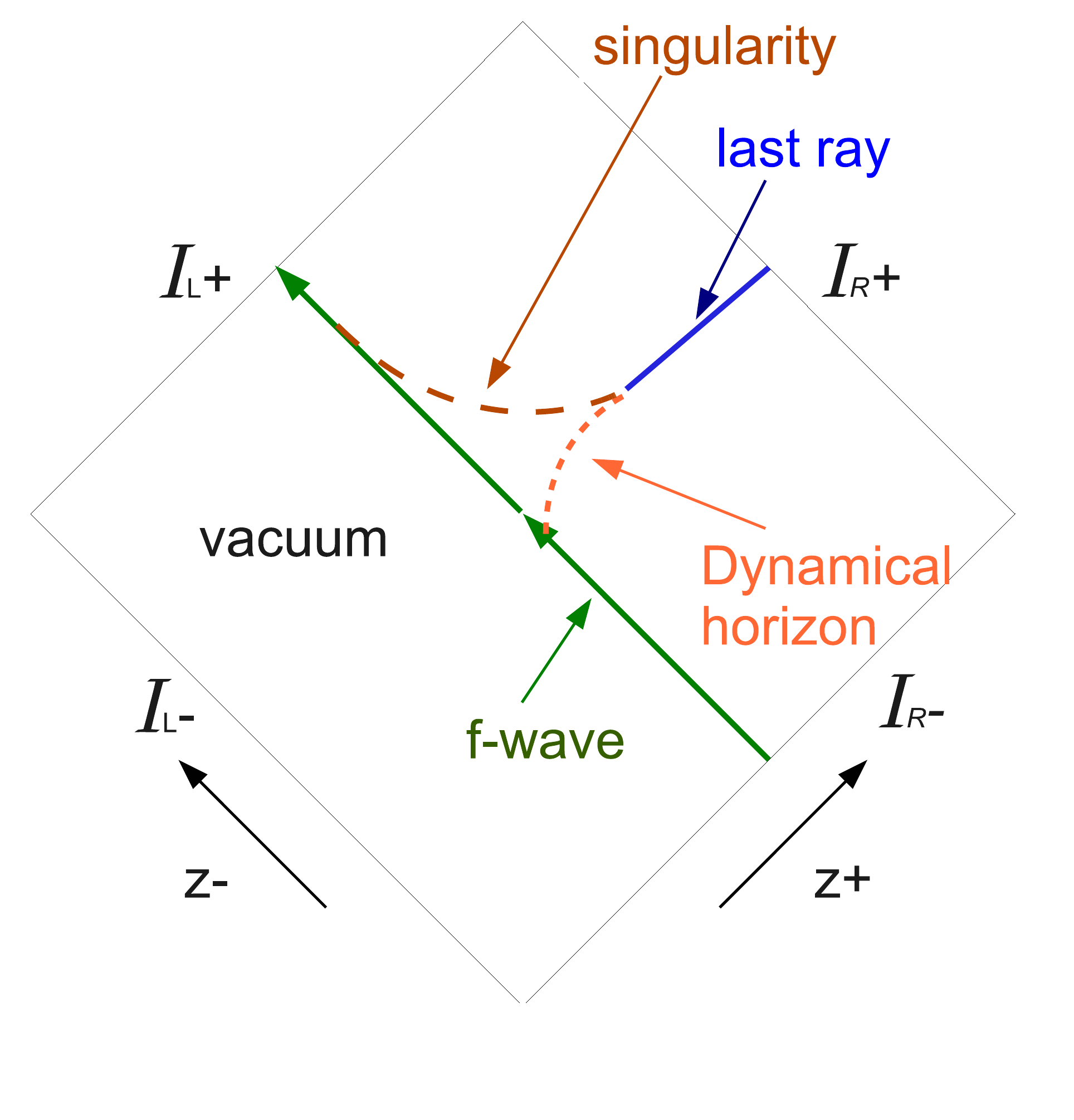}\\ {(b)}
           \end{center}
    \end{minipage}
\caption{\footnotesize{ (a)  Classical collapse of a narrow pulse of a massless scalar field, incident from $\scrim$. Note that during the collapse a dynamical horizon (labelled DH) is formed. It is space-like and grows in area as the collapse continues and smoothly joins on to the (null) event horizon (labelled EH) in the distant future.\\
 (b)  Artist's depiction of  a semi-classical Callen-Giddings-Harvey-Strominger (CGHS) black hole that was analyzed in detail in \cite{aprlett,apr}. A Delta-distribution pulse of a scalar field from $\scrim$ would lead to a black hole in the classical theory. In the semi-classical theory, the singularity is weakened (the metric is $C^{0}$ but not $C^{1}$) but persists.  The dynamical horizon becomes time-like, shrinks in its `area' due to emission of quantum radiation, and  terminates and meets the singularity. The dynamical horizon together with the last ray from its end point to $\scrip$ constitutes the future boundary of the semi-classical space-time.}}
 \label{fig:3}
\end{center}
\end{figure}

In this section we will expand on the three themes that lie at the heart of the LQG-based paradigm sketched above.  Before doing so, however,  we need to {clarify a preliminary issue that is either implicit or overlooked in most discussions.} In Fig 1, the black hole is initially formed by the gravitational collapse of a star.  However, the star is generally treated as an ``external matter field'' in that its constituents do not feature in the black hole evaporation process. Indeed, in Hawking's original analysis, the black hole evaporates  through emission of quanta of a massless, test scalar field, a field that does not feature in the composition of the star that collapses. Therefore, we do not have a closed system. 

However, discussion of unitarity and potential loss of information in the quantum theory requires a closed system.  The simplest way to construct such a system to consider only a massless scalar field that is coupled to gravity. Thus, the black hole itself is formed by the collapse of a massless scalar field that is incident from $\scrim$ as in Fig. 3(a), rather than by a star collapse as in Fig. 1(a). More precisely, in classical GR the incoming state is a narrow pulse of a scalar field --depicted by the shaded null strip--  incident from $\scrim$ that collapses and forms a black hole. The pulse is chosen to be narrow so we have a prompt collapse and analysis is not excessively contaminated by the details of the pulse profile. It is also chosen to have a `macroscopic' Arnowitt-Deser-Misner (ADM) mass to ensure that classical GR is an excellent approximation in the early stage of the collapse. To the past of the pulse space-time is flat and, to the future of the pulse, it is isometric to a Schwarzschild space-time. (Sometimes, it is convenient to idealize the situation and consider a Delta-distribution for the pulse profile as in Fig. 3(b). ) In the quantum theory, one is interested in the Hawking radiation of \emph{this} scalar field. Thus we now have a closed system. 

More precisely, in place of the star that originates from past time-like infinity  \emph{and} the incoming vacuum state of a massless test scalar field on $\scrim$ that Hawking considered, in quantum theory we would only have a coherent state of the quantum, massless scalar field, peaked at the classical pulse, and we approximate the collapse using semi-classical gravity until curvature becomes Planckian.  Since the quantum fluctuations in a coherent state at $\scrim$  are the same as those of the vacuum state, they would again lead to the Hawking process until the black hole shrinks to a sufficiently small mass and quantum gravity proper is needed. 

There is, however, one important consideration that has to be kept in mind in the discussion of entanglement. Consider the hypersurface $\Sigma$ depicted in Fig. 2(a) during the semi-classical phase of the evaporation process. The outgoing modes that go to $\scrip$ `lie on the portion of $\Sigma$' that is to the right of  the time-like part of \emph{T-DH}. These are entangled with the entire state that `lies on the left portion of $\Sigma$' which includes both, modes carrying negative energy flux, \emph{and the ingoing coherent state}. In the literature one often considers entanglement of the outgoing modes only with the negative energy ingoing modes following Hawking's original calculation in which the collapsing star/field are is included in quantum considerations.

\subsection{Dynamical horizons}
\label{s2.1}

Let us begin by recalling the notion of the event horizon (\emph{EH}) of a black hole.  \emph{EH} is the future boundary of the space-time region from which causal signals can reach $\scrip$. Note that this notion is not only extremely global, but also \emph{teleological.} Thus, for example, an \emph{EH} may well be developing in the room you are now sitting \emph{in anticipation} of a gravitational collapse that may occur in this region of the galaxy a million years from now! Indeed, Fig. 3(a)  provides a concrete illustration of how this can happen: the \emph{EH} \emph{ forms and grows in the flat part of space-time} in anticipation of the scalar field collapse that occurs later. Furthermore, as discussed in a remark  below, in models where back reaction has been analyzed in detail, there is no \emph{EH} in the semi-classical part of the space-time. And it will not materialize in the quantum completion of the space-time \emph{if} the singularity is resolved. What do we mean by `black hole evaporation' then? Something non-trivial does happen already in the semi-classical part of space-time: As mentioned in Section \ref{s1}, what forms and evaporates is a dynamical horizon, \emph{DH}.

A \DH  is a 3-dimensional space-like or time-like submanifold that is: (i)  foliated by  2-dimensional, surfaces $S$ with 2-sphere topology, so that the expansion of one of the null normals to each leaf $S$ is zero and that of the other null normal is either positive or negative everywhere.%
\footnote{For brevity, in what follows I have skipped finer points, in particular implicitly assuming spherical symmetry at a few places lateron. I have also simplified the terminology a little bit from the original work \cite{ak1,ak2,bbgvdb}, whose primary focus was on classical GR.  For reviews, see \cite{akrev,boothrev}.} 
 Thus, each $S$ is a marginally trapped surface (MTS).  In an asymptotically flat space-time,  we can distinguish between the two null normals to $S$ -- we will denote by $\l^{a}$ the outgoing null normal and by $n^{a}$ the ingoing null normal. On a BH type \emph{DH}, the expansion $\Theta_{(\ell)}$ of the outgoing null normal vanishes (it is positive immediately outside and negative immediately inside the MTS), while the expansion $\Theta_{(n)}$ of the ingoing null normal is negative (both outside and inside). Thus, immediately inside a black-hole type \emph{DH}, both expansions are negative and we have a trapped region. Therefore, we will refer to these \DHs as  \emph{trapping dynamical horizons,} \emph{T-DHs}. One can show that the \TDH is space-like when the area of the MTS increases along the projection of $\l^{a}$ on \emph{DH}, i.e. in the outward direction.  More importantly, there is an explicit, precise relation between the growth of the area and the flux of  energy (carried by matter and/or gravitational waves) flowing \emph{into} the horizon: If $R$ is the area radius of MTSs, then \cite{ak1,ak2}
 \be \label{arealaw}  \f{R_{2}}{2G} - \f{R_{1}}{2G} = \hbox{\rm Energy flux into the portion $\Delta$(\emph{T-DH})}\, , \ee 
where the MTS with area radius $R_{2}$ is outward relative to that with area radius  $R_{1}$ and  $\Delta$(\emph{T-DH}) is the portion of the \TDH bounded by these two MTSs. Thus, not only does the second law of black hole mechanics hold for every  \TDH but the growth of the horizon area is \emph{directly} related to the physical process of energy infall. This is in striking contrast with the situation for \emph{EHs},  where we only have a qualitative statement of growth. Indeed, it is not possible to directly trace the growth to the infall of energy locally because, as we see in Fig 1(a) \emph{EHs} can form and \emph{grow} in flat space-time where there is nothing at all falling across it.

During the evaporation process, by contrast, the MTSs on the  \TDH shrink, again following (\ref{arealaw}), and the \TDH is now time-like.  
%
%
 In the CGHS black hole\, --depicted in Fig. 3(b)-- \,that was analyzed in detail using highly accurate numerical simulations, not only does the \TDH exhibit all these features both during the collapse and evaporation, but it has additional properties that have direct physical interpretation. Let me mention two. First,  the Hawking radiation starts (in earnest) precisely at the (retarded) instant of time at which the \TDH  has maximum area. Second, as the evaporation proceeds, the area of the \TDH at any time instant is directly related to the Bondi mass at the corresponding retarded time instant at $\scrip$ \cite{ori2,aprlett,apr,ori3}.

Finally, let us consider white-hole type \emph{DHs}. Now, it is the expansion $\Theta_{(n)}$ of the \emph{ingoing} null normal that vanishes and the expansion $\Theta_{(\ell)}$ of the outgoing null normal is positive. Thus, immediately inside these horizons, both expansions are positive: we have an anti-trapped region. We will therefore refer to these horizons as \emph{anti-trapping dynamical horizons} \emph{AT-DH}.  Since it is $\Theta_{(n)}$ that vanishes on any \emph{AT-DH}, it is natural to investigate what happens to the area of the MTSs as one moves along the projection of $n^{a}$ on the \emph{AT-DH}. If the \ATDH is space-like, its area \emph{decreases}  (now in the inward direction) and if it is time-like its area increases (now in the future direction).

Thus main differences between \emph{EHs} and \DHs are the following. First,  \emph{EHs} refer to  the global structure of space-time and are teleological, while \DHs can be located quasi-locally and their properties have direct relation to physical processes at their location. Second, \emph{EHs} are null while \DHs can be space-like or time-like, and become null only when they become `isolated' i.e. there is no flux of energy across them. Third,  nothing can \emph{ever} escape to the `exterior' region from the trapped region enclosed by a BH-type \emph{EH} and nothing can \emph{ever} enter the anti-trapped region bounded by a WH-type \emph{EH}. 
While there are trapped surfaces immediately inside a \emph{T-DH},  one can send causal signals across a \TDH from inside to outside (see  Figs. 2(b) and 4(b)). Similarly, there are future directed causal curves that traverse an  \ATDH from outside to inside.\\

\emph{Remark:}  As emphasized by Geroch and Horowitz \cite{rggh},  for the notion of an \emph{EH} to be useful in discussions of black holes, it is essential that $\scrip$ be complete; otherwise even Minkowski space could have an \emph{EH}! I mention this caveat explicitly because it is often overlooked in the discussion of information loss. Consider the gravitational collapse of a shell depicted in Figs. 3.  In the classical theory, there is indeed an  \emph{EH}  formed by the  (thick) shell collapse because $\scrip$ is  complete and the \emph{EH} is the future boundary of the causal past of $\scrip$ (see Fig. 3(a)).  However, the situation changes when back reaction is included.  The evaporation process has been studied in detail for the CGHS black hole in the semi-classical regime (see in particular \cite{ori2,aprlett,apr,ori3})  and  concrete ideas have been put forward for how the semi-classical discussion can be completed in quantum gravity \cite{atv}. This evaporation process of is depicted in Fig. 3(b).  The singularity persists also in the semi-classical space-time (although it is milder in that the space-time metric is now continuous across the singularity). Therefore the semi-classical space-time has a future boundary depicted by the `last ray' in Fig. 3(b). It is tempting to interpret it as the \emph{EH}. However, one can check explicitly that $\scrip$ of this semi-classical space-time is incomplete and therefore the last ray is \emph{not} a portion of an \emph{EH} \cite{apr}.%
\footnote{In fact, in this specific case, space-time metric is completely regular across the last ray; there is no `thunderbolt' suggested by Hawking and Stewart \cite{swhjs}.  Therefore, 
in a more complete, quantum theory, one can expect space-time to extend further. If the singularity were to be resolved in full quantum gravity, then the quantum space-time would have a longer $\scrip$  as suggested in Fig 3(b). General conditions under which unitarity can be restored in this space-time are discussed in \cite{atv}.}
Indeed,  the semi-classical part of the quantum space-time has no event horizon at all: what forms and evaporates in this space-time is a  \emph{T-DH}.

\subsection{No violation of semi-classical expectations}
\label{s2.2}

Consider the  space-time depicted in Fig.  2(a) governed by the coupled semi-classical gravity equations
\be \label{semi-class} G_{ab} = 8\pi G_{\rm N}\,  \langle\, \h{T}_{ab}\, \rangle \quad {\rm and} \quad  \Box\, \h{\Phi} = 0 \ee
and the state of $\h\Phi$ depicting  a prompt collapse of a (thick) shell of the scalar field incident  from $\scrim$, depicted by the shaded region in the figure. The collapse gives rise to the formation  a space-like \emph{T-DH}, shown by the left part of the (dark) line labelled \TDH in the support of the collapsing scalar field.  This part starts with zero area (at the top left corner) and the area of MTSs continues to grow rapidly (in the outward direction) as long as there is incoming flux of (positive) energy carried by the scalar field. If we were to `switch-off' quantum effects, then the situation would be as in Fig. 3(a):  Once the incoming energy flux terminates, this \TDH would smoothly join on to a null isolated horizon --the future part of  \emph{EH}.  The singularity would stretch all the way to $i^{+}$. However, because of quantum effects, the fluctuations in the incoming coherent state give rise to spontaneous (and stimulated) emission. Modes are created in pairs, one going in, and the other going out to $\scrip$. There is thus a flux of \emph{negative} energy into the `interior region' whence, in contrast to the classical theory,  the \TDH does not become a null isolated  horizon. Instead, it becomes null only instantaneously and continues as a time-like 3-surface as depicted in the right branch of \TDH in Fig. 2(a).  As the negative energy flux across this portion of the \TDH continues, the area of MTSs on this portion of the \TDH decreases.  Recall, however, that by assumption the ADM mass of the system is `macroscopic' --say a solar mass, $M_{\odot}$. Then initially  the ingoing flux is \emph{very} small and, correspondingly, the area also decreases \emph{very} slowly. Indeed, the radius starts out at $\sim\, 3$km  (corresponding to the dynamical horizon mass $M_{\TDH}= M_{\odot}$) and to reach, say, $\sim\,  0.1$mm (so that the $M_{\TDH}$ now equals lunar mass, $M_{\rm L}$),  it takes some $10^{64}$ years!  

For concreteness, in this subsection \emph{we will focus on this phase of evaporation, during which a solar mass $\TDH$ shrinks to a lunar mass \emph{T-DH}. } Now, during this long process there is no a priori  reason to expect a violation of semi-classical considerations because we are \emph{very} far from the Planck regime throughout this slow transition. However, this premise does violate an intuition based on considerations of the `page time \cite{page} that many in the community rely on. Our strategy is to assume that semi-classical considerations are valid for reasons just mentioned, and see if one runs into roadblocks that force us to concede that the assumption is incorrect. 

Calculations have been carried out in some detail  within the semi-classical framework to support the scenario I summarize and I believe that it is possible to make the entire analysis rigorous (say by Phys. Rev. D  or J. Math. Phys. standards)  using quantum field theory in curved space-times  to analyze in greater detail the behavior of partner modes in the dynamical background space-time. Pairs of Hawking modes are created continually during this long process, one going to infinity and its partner falling inside the time-like portion of the \emph{T-DH}.  These modes will be entangled whence, if one uses the usual observable algebra based just at $\scrip$,  the state would seem mixed, close to a thermal state. (For caveats on departure from thermality, see, e.g.  \cite{anderson}).  When will the correlations be restored? For this to happen, the partner modes would have to emerge from the trapped region and propagate to $\scrip$.  

Now,  one's first reaction could be that this `purification of the state' could occur continually throughout the  $10^{64}$-years long process because there is no causal obstruction for the modes that entered the trapped region bounded by the  \TDH  to come out across its long  time-like portion.  Indeed arguments have been made to say that there is no information loss issue at all because there is no event horizon and fields can escape across the time-like 
\TDH (see, e.g., \cite{hayward-pop,hayward-conf}). But in  semi-classical gravity, throughout this process partners of modes that go out to $\scrip$ continue to fall into the trapped region; they do not come out. So the fact that there is no causal obstruction for the partner modes to emerge from the trapped region does not by itself lead to purification in the semi-classical space-time.

Furthermore, there is an important puzzle. During the long process, a `very large number of modes' has escaped to infinity (some $10^{75}$ by a back of the envelope heuristic counting). But the  horizon mass $M_{\TDH}$ is now the lunar mass, $M_{\rm L}$, only about $10^{-7}$ times $M_{\odot}$.  In terms of modes then,  those that escaped to infinity have carried away almost all the initial $M_{\odot}$ and their partner modes have only $\,10^{-7} M_{\odot}\,$.  Therefore, at the end of the semi-classical process under consideration, one would have to have a huge number $\mathcal{N}$ of modes carrying   a total of $M_{\odot}$ in energy  to $\scrip$, and the same number $\mathcal{N}$ of modes inside the \TDH with a minuscule mass. How can a \TDH with just a $0.1$mm radius accommodate all these modes? Even if we allowed each mode to have the (apparently maximum) wavelength of $0.1$mm, heuristically one would need the horizon to have a huge mass --some $10^{22}$ times the lunar mass!  While these considerations are quite heuristic and the semi-classical calculations are much more precise, one needs to face the conceptual tension: At the end of the process under consideration, the \TDH has simply too many modes to accommodate, with a tiny energy budget. Is this not an indication that, contrary to what semi-classical considerations suggest,  most of the `purification' must have occurred by the time the \TDH has shrunk down to the lunar mass? That, as is often argued,  `purification' must begin already by Page time  \cite{page} when the \TDH has lost only half its original mass of $M_{\odot}$? If so, semi-classical considerations must fail even for macroscopic black holes; there must be unforeseen quantum effects to  get us out of this quandary.

This does seem like a serious problem with the original assumption on validity of semi-classical theory for the process under consideration till one takes a more careful look at the space-time geometry in the trapped region. It turns out that, because of semi-classical Einstein's equations,  this geometry has some very counter-intuitive features that can resolve this quandary. Calculations of the stress-energy tensor on the Schwarzschild space-times confirm the idea that, in semi-classical gravity there is a negative energy flux across the time-like portion of $\TDH$ such that $M_{\TDH}$ would decrease according to the standard Hawking formula: $\rmd M_{\TDH}/\rmd v = -  \hbar/(GM_{\TDH})^{2}$. Indeed, this is the basis of the standard view that the evaporation time goes as $\sim M^{3}_{\rm ADM}$. One can then argue that, in the phase of evaporation under consideration,  the form of the space-time metric in the region bounded by the  $\TDH$ of Fig 2(a) is well approximated by  the Vaidya metric:
\be \label{metric} {\rmd} s^{2}  = - \big(1- \f{2Gm(v)}{r} \big){\rmd} v^{2} + 2 {\rmd} v {\rmd} r + r^{2} \big({\rmd}\theta^{2} + \sin^{2}\theta\, {\rmd} \varphi^{2}\big), \ee
with $m(v) = M_{\TDH}(v)$.  (This is because during this phase the quantum correction to the Schwarzschild metric of classical GR are small.)  In particular, analogous conclusion is borne out in the detailed semi-classical analysis of evaporation of the CGHS black hole. So, let us work with this metric. To understand the nature of space-time geometry it bestows on the trapped region, it is convenient to foliate it by some invariantly defined surfaces.  It turns out that the most convenient choices are \cite{ao}:  $r = {\rm const}$ and $\Kr = {\rm const}$ where $r$ is the area radius of the round 2-spheres and $\Kr$ is the Kretschmann scalar: 
\be \Kr \, :=\, R_{abcd} R^{abcd}\, =\, 48\,  \f{G^{2}m^{2}(v)}{r^{6}}, \ee
where the last equality expresses the value of the Kretschmann scalar in the Vaidya metric. (Note that it has the same form as in the Schwarzschild space-time, with the constant $m$ of the Schwarzschild space-time replaced by $m(v)$; derivatives of $m(v)$ do not appear!)  A typical leaf of these foliations is represented in Figs. 2 by the portion of the Cauchy surface $\Sigma$ that lies in the trapped region, the left end of the leaf lying on the space-like portion of \TDH and the right end on the time-like portion.

If we were to ignore the quantum radiation as in Fig. 3(a) depicting the scalar field collapse  in classical GR,  $m(v)$ would be constant, the metric in the trapped region would be just the Schwarzschild metric and both foliations would coincide, each leaf being the standard homogeneous slice in the Schwarzschild-interior. However, because of quantum radiation $m(v)$ is a monotonically decreasing function of $v$ and the two foliations are distinct. If we consider the \emph{full} trapped region (beyond the semi-classical part discussed in this sub-section),
the $\Kr = {\rm const}$ foliation is conceptually better suited to separate the transition between the part where semi-classical considerations provide an excellent approximation, and the part where curvature enters the Planck regime and corrections from full quantum gravity become crucial. However, while each leaf is space-like almost everywhere starting form the left end,  it becomes null \emph{extremely} close to the right end.  Thus each leaf has a time-like portion, albeit \emph{extremely} tiny. The $r={\rm const}$ foliation on the other hand is space-like everywhere and easier to visualize. However, this foliation does not cleanly separate the trapped region into a part on which semi-classical considerations provide an excellent approximation and part where corrections from quantum gravity proper become important.  Thus, each has its merits and it is appropriate to consider both these foliations. Both provide a foliation of the entire trapped region of Fig. 2(a).%
%
\footnote{Another geometrically natural family of 3-surfaces is defined by constancy of the trace of the extrinsic curvature is constant \cite{cdl}. But they do not provide a foliation of the entire trapped region because if we start from the right end --the time-like piece of \emph{T-DH}-- the leaves `pile up', all approaching the 2-sphere $r= 1.5 M_{\odot}$ on the left end. But the main phenomenon of developing enormously long necks occurs also on these 3-surfaces. This foliation was motivated by a calculation of the growth of the volume of 3-surfaces inside the event horizon of a collapsing shell in classical GR \cite{mccr}.}
 
Geometry of the leaves of these foliations can be analyzed simply by pulling back the space-time metric (\ref{metric}) to them. Each leaf is topologically $\mathbb{S}^{2} \times \mathbb{R}$ and is itself foliated by round 2 spheres which can be labelled by values of the advanced time coordinate $v$. Let us set $v=0$ when  $M_{\TDH} = 1M_{\odot}$, and $v=v_{0}$ when it reaches the lunar mass $M_{L}$, both at the right end, i.e., on the time-like portion of the \emph{T-DH}.  Thus, the radius of the MTSs decreases from $r|_{v=0} = 3$Km to $r|_{v=v_{0}} = 0.1$mm. In both foliations, as $v$ increases  the leaves develop longer and longer necks of length $\ell_{N}$ along the $\mathbb{R}$ directions (while the increase in the 2-sphere radius is much less pronounced).  The `final leaf' for the process under consideration in this sub-section starts at the right end with $v=v_{0}$ where $M_{\TDH}$ equals $M_{L}$.  For the two foliations under consideration, the length $\ell_{N}$ of the final leaf is given by:
\vskip0.2cm
\centerline{$\Kr = {\rm const}$  foliation: $\,\,\ell_{N} \approx 10^{64}$\,lyrs,  $\qquad\qquad$  $r={\rm const}$ foliation: $\,\, \ell_{N} \approx 10^{62}$\,lyrs;}
\vskip0.2cm
\noindent  where `lyr' stands for light years.  The radius of 2-spheres is constant, $r\sim 10^{-2}$cm on this entire leaf for the $r={\rm const}$ foliation, and increases from $\sim10^{-2}$cm to $\approx3\times 10^{5}$cm from  the right to the left end for the $\Kr = {\rm const}$ foliation. Thus, during the slow evaporation process, \emph{the leaves develop astronomically long necks!}
 
 The dynamical nature of the intrinsic geometry of these leaves suggests that the partner modes that fall across the time-like part of the \TDH will get enormously stretched during evolution from $v=0$ to $v=v_{0}$, as in quantum field theory on an  expanding cosmological space-time, and become infrared. As I mentioned, a back of the envelope calculation suggests  an enormous number -- some $10^{75}$-- of these partner modes is created during the long process during which the black hole shrinks from a solar mass to a lunar mass! But with such infrared wavelengths, it is easy to accommodate them in the trapped region with the energy budget only of $M_{\rm L}$.  Thus, even though the  outgoing modes carry away $(1- 10^{-7})\, M_{\odot}$  mass to $\scri$, there is no obstruction to housing all their partners in the trapped region on a slice $\Sigma$ of Fig. 2(a) with the small energy budget of just $10^{-7}M_{\odot}$.  This argument removes the necessity of starting purification by Page time. In the LQG perspective of this article, purification can be postponed to a much later stage.
 
Of course these considerations need to be made precise using quantum field theory on the Vaidya space-time of Eq. (\ref{metric}).  {In particular,  while the qualitative phenomenon of  the `stretching of the wave-length' is familiar from the theory of cosmological perturbations in the inflationary epoch, the process and the associated energy considerations need to be spelled out using renormalized stress-energy tensor of the full the scalar field $\h\Phi$ --which, in the trapped region, includes \emph{both} the infalling part responsible for the collapse, and  the partner modes created during the evaporation process. It is \emph{total} energy in the trapped region that diminishes as the field evolves to the future.} But this is feasible using available techniques form quantum field theory in curved space-times by exploiting spherical symmetry of the semi-classical geometry.  Finally, although the foliations discussed above are geometrically natural,  from a covariant perspective, one may have reservations about using any foliation at all. However, one can think of these foliations simply as mathematical tools that provide us intuition to help answer invariantly defined questions, as is commonly done in cosmological contexts. 
 
\subsection{Singularity resolution and the quantum region}
\label{s2.3}

The question remains, of course, as to how purification is to occur at a later time. Now, if the space-time geometry in the \emph{entire} trapped region were to be given by the Vaidya metric, a space-like singularity would constitute the future boundary of the trapped region (as in, e.g. \cite{mdcr}). Then in our scenario there would be no chance for all the partner modes to  escape to $\scrip$ and restore correlations with the modes that reached $\scrip$ during the long semi-classical phase (i.e. by the retarded time $u= u_{\rm LR}$, defined by the last ray  in Fig 2 (a)).  Thus, purification can occur only if the space-like singularity of the  Vaidya metric is resolved because the quantum geometry in the Planck regime is qualitatively different and singularity-free. Can LQG quantum geometry naturally resolve the singularity? In this subsection we will address this issue in two steps. In the first we  will recall some results on the resolution of the black hole singularity in the non-dynamical Kruskal space-time. In the second we will summarize the current expectations for singularity resolution for the dynamical space-time now under consideration, in which the trapped region results from gravitational collapse of the quantum scalar field $\h\Phi$, and there is  further quantum dynamics because of creation of the Hawking pair of modes that react back on the space-time geometry.

\subsubsection{Kruskal space-time in LQG}
\label{s2.3.1}

Recall that the big-bang singularity of Friedmann-Lemaitre-Robertson-Walker (FLRW) space-times are naturally resolved by the quantum geometry effects of LQG (see, e.g. the review articles \cite{asrev,iapsrev}). The big-bang is replaced by a big-bounce, to the past of which we have a contracting phase and to the future of which we have an expanding phase. Space-time curvature reaches an \emph{universal upper bound} at the big-bounce. Recall that  at the big-bang  the radius of space-time curvature $\r$ vanishes, while in Minkowski space-time $\r=\infty$. In loop quantum cosmology (LQC) of FLRW models, the radius of curvature has an absolute lower bound $\r \sim 0.13\lp$. Now, the trapped region of the Kruskal space-time --often called the `Schwarzschild-interior' in the literature-- is isometric with a Kantowski-Sachs metric, which is foliated by homogenous slices, each with topology $\mathbb{S}^{2}\times\mathbb{R}$. Therefore,  over the years LQC techniques were applied to the interior to probe the issue of singularity resolution (see, e.g.,  \cite{ab,lm,bv,dc,cgp,bkd,cs,djs,oss,cctr,abp,nbetal}). However, because  the Kantowski-Sachs space-time is anisotropic --and furthermore anisotropies are technically different from those of the better understood Bianchi models \cite{asrev,iapsrev}-- progress has been slower. While the LQC quantum equations can be written down also for the Schwarzschild-interior, only the effective equations have been solved in detail. These provide a quantum corrected  \emph{effective metric} for the Schwarzschild-interior --a smooth tensor field that shares the isometries of the Schwarzschild metric but with coefficients that involve the Planck length $\lp$. 
The black hole singularity is resolved in all investigations  but details of quantum modifications differ. For brevity, I will restrict myself to one \cite{aoslett} which is free of all the known drawbacks  (for a detailed comparison of quantum modifications, see \cite{aos}).

\begin{figure} 
  \begin{center}
    \begin{minipage}{2.5in}
      \begin{center}
        \includegraphics[width=2in,height=2.2in]{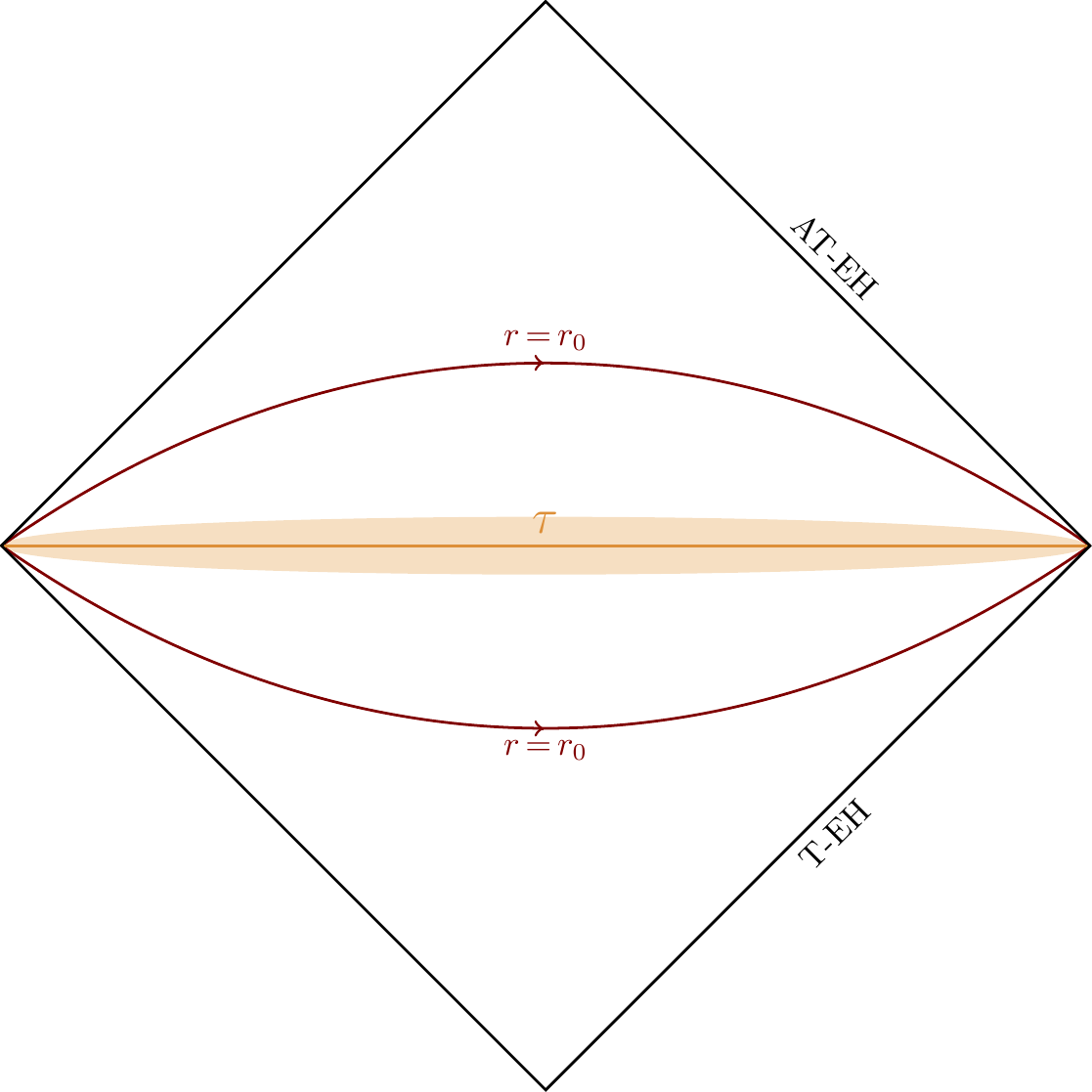}\\ {(a)}
                 \end{center}
    \end{minipage}
   \hspace{.5in}
    \begin{minipage}{3in}
      \begin{center}
       \includegraphics[width=2.3in,height=2.7in]{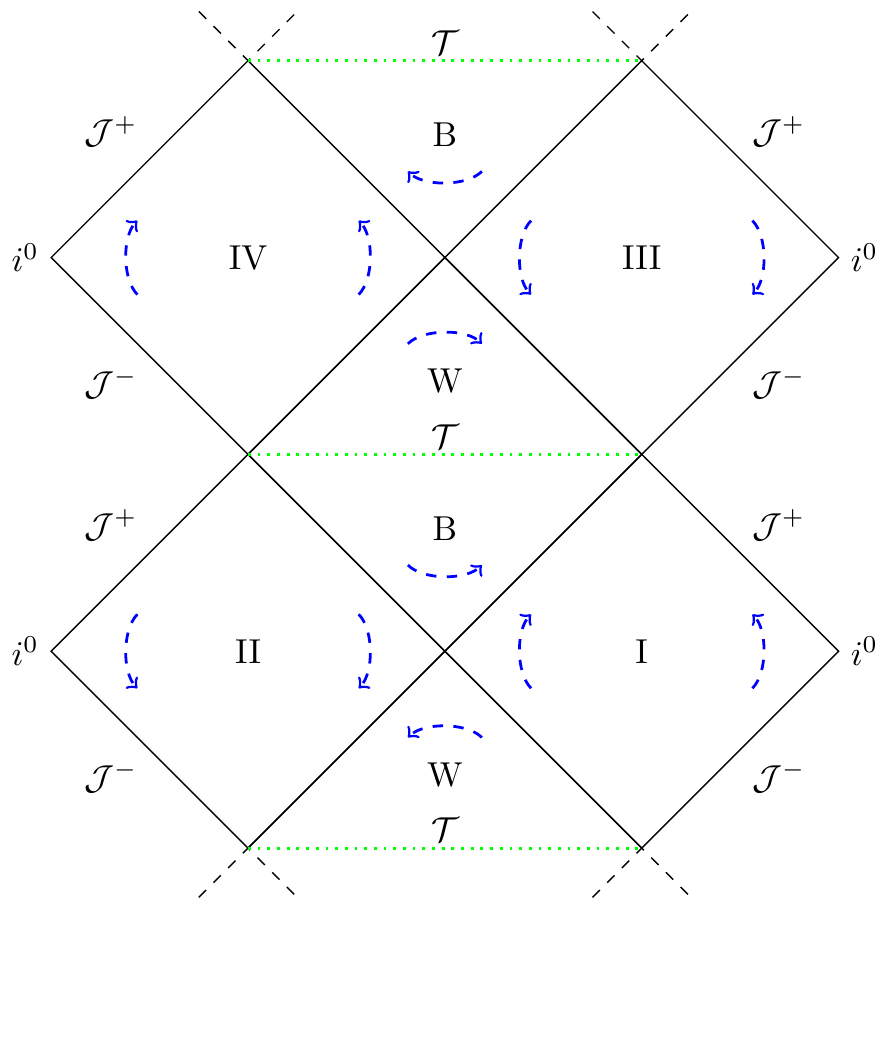}\\ {(b)}
              \end{center}
    \end{minipage}
\caption{\footnotesize{(a)  LQG extension of the Schwarzschild-interior. In classical GR we only have the lower triangular region that is bounded below by the trapping event horizon \emph{T-EH} and ends with a future space-like singularity at $r=0$. Quantum geometry effects of LQG resolve this singularity and replace it with a \emph{Transition Surface}, labelled $\tau$, to the future of which we have an anti-trapped region with a future boundary that constitutes an anti-trapping event horizon \emph{AT-EH}.  Quantum geometry effects are important only in a neighborhood of $\tau$, shown by a shaded (pink) region.\\
(b) LQG extension of the Kruskal space-time \cite{aoslett,aos}. Quantum geometry effects resolve both black hole and white hole singularities and the quantum corrected LQG space-time extends indefinitely.  The central diamond corresponds to Fig. (a). The dashed (blue) arrows depict integral curves of the Killing vector. Just as in only a part of the full Kruskal space-time is physically relevant for gravitational collapse in classical GR, only a portion of this infinite extension is relevant for the black hole evaporation process (see Fig. 2(b)).}}
\label{fig:4}
\end{center}
\end{figure}

As Penrose has emphasized over the years, there are key structural differences between the FLRW and Schwarzschild singularities in classical GR because in the first case it is the Ricci curvature that diverges while in the second case it is the Weyl curvature.  {While there are `normal'  2-spheres arbitrarily close to the big-bang (with $\Theta_{(\ell)} \cdot \Theta_{(n)} <0$), every 2-surface in the region near the Schwarzschild singularity is trapped (with $\Theta_{(\ell)} \cdot \Theta_{(n)} >0$).} Therefore, there are some important differences also in the quantum theory. However, rather unexpectedly, there are also close similarities. I will present a few illustrations while discussing the singularity resolution in the Schwarzschild-interior.

The main features of the singularity resolution can be summarized as follows.  As depicted in Fig. 4(a),  the Schwarzschild singularity is replaced by a (space-like) transition surface $\tau$ in LQG. To the past of $\tau$ we have a trapped region, expansions $\Theta_{(\ell)}$ and $\Theta_{(n)}$ of the two null normals $\ell^{a}$ and $n^{a}$ to the round 2-spheres are both negative. Both expansions vanish on $\tau$ and become positive to the future of $\tau$. Thus, we have an \emph{anti-trapped} region to the future. The radii of round 2-spheres decrease as we approach $\tau$ from below and increase as we move upwards. Therefore the radius attains its minimum value at $\tau$. This is analogous to the fact that the scale factor $a$ attains its minimum value at the LQC big bounce in FLRW models.%
\footnote{If the spatial topology is non-compact,  the scale factor $a$ does not have an invariant meaning. In this case, the spatial region with a fixed comoving volume has the minimum \emph{physical} volume at the LC bounce.} 
Also, curvature attains its maximum value at the transition surface, and that value is universal (for macroscopic ADM masses assumed in obtaining effective equations). For example, the Kretschmann scalar is given by:
\be
\Kr \mid_{\tau}\,\, =\,\, \f{\alpha\,\, \lp^{8}}{\Delta^{6} }+{\cal O}\Big(\big( \f{\lp}{GM}\big)^\f{2}{3} \,\ln\f{GM}{\lp}\Big) \lp^{-4}\ee
where $\alpha$ is a numerical constant and $\Delta \sim 5.17\, \lp^{2}$ is the area gap of LQG, and $M$ the ADM mass of the Schwarzschild space-time under consideration. This corresponds to a \emph{universal minimum,} $0.057 \lp$, for the curvature of radius $\r$  in the LQG extension of the Schwarzschild-interior. In the FLRW model, curvature attains an universal upper bound at the bounce surface and the corresponding lower bound for the curvature radius is $\r^{\rm FLRW} = 0.127 \lp$. 
Thus, rater surprisingly, the minimum values are the same within a factor of $\sim 2$ although the structure of singularities and the detailed analyses leading to these values are very different. Next, in the LQC of FLRW models quantum geometry corrections decay rapidly away from the bounce. The same is true in the Schwarzschild-interior, as one moves away from the transition surface $\tau$. For example, even in the case when $M_{\rm ADM} = 10^{6} \mp$, the quantum geometry corrections at the horizon are less than 1 part in $10^{8}$. So, the geometry is well approximated by the Schwarzschild solution till one reaches a small neighborhood of $\tau$. Thus, even a black hole with mass as low as $10^{6} \mp$ qualifies as a `macroscopic' in this effective theory.

A major difference between FLRW models and the Schwarzschild-interior is  that in the classical theory it is the presence of matter that gives rise to non-trivial curvature in the former, while there is no matter in the latter. In both cases singularity is resolved because the left hand side --i.e. the geometrical part-- of Einstein's equations receive corrections due to LQG quantum geometry. In the FLRW case, one can manipulate the effective equations and  interpret the new terms as corrections to the matter part in the Friedmann and Raychaudhuri equations. Then the LQC corrected Friedmann equation, for example, reads \cite{apslett}
 \be \Big(\f{\dot{a}}{a}\Big)^{2} = \f{8\pi G}{3} \, \rho\, \Big(1 - \f{\rho}{\rho_{\rm sup}} \Big) \qquad {\rm where}\qquad  \rho_{\rm sup} = \f{\beta}{\Delta^{3}} \, \approx\,   0.41\, \rho_{\rm Pl}, \ee
  $\beta$ being a numerical constant. Thus the area gap  is the \emph{microscopic} LQG parameter that dictates the \emph{macroscopic} parameter $\rho_{\rm sup}$ which determines the regime in which LQG effects are important. The negative sign in front of the quantum correction $\rho/\rho_{\rm sup}$ implies that effective energy density is lower than in the classical theory, but the negative quantum contribution becomes quickly negligible  away from the Planck regime.  In the Planck regime however, it creates a very large effective repulsive force  that is responsible for the resolution of the big-bang singularity. In the case of Schwarzschild-interior the mechanism is similar and can be seen even more directly. In the classical theory, $R_{ab} =0$ and curvature resides entirely in the Weyl part. But because of quantum geometry effects, $R_{ab}$ is no longer zero in the effective theory and  one can choose to interpret it as an effective stress-energy tensor induced by quantum geometry. For macroscopic black holes one finds that the invariant  $R_{ab}R^{ab}$ is negligible near the horizon but increases very rapidly near $\tau$, reaching the Planck scale value $ (\alpha\,\, \lp^{8})/(3\Delta^{6})$ at $\tau$, and then decreases rapidly again in the anti-trapped region.  The effective stress energy tensor $\mathcal{T}_{ab}$ has the form of the stress-energy tensor of a perfect fluid. As one would expect, the corresponding energy density and pressure are negative. Just as the negative sign in front of the quantum correction to the Friedmann equations arises naturally from the LQG quantum geometry without additional inputs, the negative sign of the effective energy density and pressure arise naturally.  They are responsible for an effective repulsive force that  leads to the singularity resolution.  It is again quite surprising that while the negative energy density becomes sufficiently high in the Planck regime to avoid a curvature blow up, it dies very quickly away from the Planck regime. Fig. 4(b) shows the LQG extension of full  Kruskal space-time \cite{aos}. It is included only for completeness. Just as the full Kruskal space-time is not directly relevant to a black hole formed by gravitational collapse or to discuss its evaporation, only a part of this picture is relevant for our present discussion. The relevant part is shown in Fig. 2(b) which we now discuss.

\subsubsection{Beyond the semi-classical region}
\label{s2.3.2}

When do quantum gravity effects become important? Consider as an example the closed FLRW model. In this case, the notion of the volume of the  universe is unambiguous.  Sometimes it is said that quantum effects become important only when the universe contracts to the Planck size, i.e. when its volume is given by $V \sim \lp^{3}$. However, detailed calculations have shown that this is not the case: What matters is when the curvature reaches the Planck size; not the volume \cite{apsv}.  In a solution in which the universe grows to a maximum radius of a megaparsec, for example, a bounce occurs when the volume of the universe is $10^{116} \lp^{3}$. Thus, in this solution quantum gravity effects completely dominate the dynamics when the universe is the size of a small mountain, and it never gets smaller. This is because in LQC the curvature reaches its maximum allowed value  already at this stage. The situation is the same in the case of a black hole: what matters is the curvature.  Consider the shell collapse, depicted in Fig. 3(a).  For a solar mass 
Schwarzschild black hole, curvature becomes Planckian already when the shell reaches radius $r\approx 1.26 \times 10^{20}\lp$, whence quantum gravity effects can dominate the dynamics already at this stage.

In semi-classical considerations of section \ref{s2.2}, we restricted ourselves to a time interval $(v=0, v=v_{0})$, during which the mass of the \TDH decreases from the solar mass, $M_{\odot}$, to the lunar mass, $M_{L}$, to first bring out and then resolve an apparent problem already in this very tame regime. But one would expect the domain of semi-classical approximation  to be much larger, say till the space-time curvature becomes $\mathcal{O}(10^{-6})\, \lp^{-2}$.  Let us then consider the corresponding $\Kr = \rm{const}$ surface within the trapped region. At the left end of this surface we encounter the infalling shell whose radius has shrunk to  $r|_{v=0} \approx 5\times 10^{14} \lp$.  At the right end of this surface, i.e., on the time-like portion of  \emph{T-DH}, we would  have  $r|_{v=v_{0}}\approx  1.8\times 10^{3} \lp$, so that the corresponding dynamical horizon mass is $M_{\DH} \approx 9\times 10^{2} \mp$. This 3-surface --depicted by the past boundary of the shaded region around the transition surface $\tau$ of Fig. 2(b)--  is the  future boundary of the semi-classical part of the trapped region. We will denote it by $\Ssc$.  In the trapped region, we  can trust the semi-classical approximation of Eq. (\ref{semi-class}) to the past of $\Ssc$  but not to its future.
 
To discuss the space-time and scalar field dynamics to the future of $\Ssc$, we need to consider 
interaction between the quantum scalar field $\h\Phi$ and \emph{quantum} geometry.  In Fig. 2(b),  the portion of space-time in which one has to go beyond semi-classical approximation is  depicted by the (pink) shaded region around the transition surface $\tau$ (together with a (red) blob at the right end).  Let us divide the problem into two parts by slicing this region by $v={\rm const}$ 3-surfaces, so that we have $m(v) = M_{\odot}$  at the left end, which then decreases as we move to the right. Since the negative energy flux is very small over a very large time span, one would expect the quantum geometry to change very slowly during this phase. As discussed below, we should be able to describe dynamics during this phase using certain adiabatic approximations.  At the end of this adiabatic phase, however, one would need full quantum gravity. Region where this will be necessary is depicted in Fig. 2(b) by the (red) blob at the right end of the transition surface $\tau$.%
\footnote{Illustrative numbers: The expectation is that one can use the adiabatic approximation discussed below during the long process in which $m(v)$ decreases from $M_{\odot}$ to, say,  $100\mp$ (when $\dot{m}(v) \sim 10^{-4}$), and the full quantum treatment would be necessary only at the every end when $m(v)$ decreases further.}

To describe dynamics during adiabatic phase we can seek guidance from other areas of physics.  Consider propagation of light --the quantum Maxwell field-- in a medium,  say water.  The medium itself is in a quantum state  $\Psi_{0}$ of a large number of molecules, and, at a fundamental level, we need to consider the interaction of photons with these molecules.  However, so long as the photon beam is not so strong as to alter $\Psi_{0}$ significantly --they do not, e.g., boil the medium--  the photons do not see the details of quantum fluctuations of molecules. The propagation can be well-approximated by computing just a couple of numbers  --the refractive index and the birefringence-- from $\Psi_{0}$  and then characterizing medium  just by these two parameters.  

The same strategy has been used in LQC. For concreteness, ler us consider the  the tensor modes of cosmological perturbations, represented by quantum fields $\h{T}_{I}$, (with $I=1,2$). At a fundamental level, they  propagate on the background FLRW \emph{quantum geometry}, described by a wave function $\Psi_{0}(a,\phi)$ where $\phi$ is the inflaton. That is $\hat{T}_{I}$ now evolve on a \emph{probability distribution} of smooth classical FLRW metric $g_{ab}$, rather than on a single $g_{ab}$.  But it turns out that so long as the back reaction of these perturbations is negligible, their propagation on the quantum geometry $\Psi_{0}$ is extremely well approximated by their propagation on a smooth metric $\t{g}_{ab}$, constructed from $\Psi_{0}$ \cite{aan3}.  $\t{g}_{ab}$ is called the \emph{dressed effective metric.}%
\footnote{Recently it was pointed out that  there is a subtle infrared problem in LQC which, however, can be overcome by restricting oneself to sharply peaked states $\Psi_{0}^{\rm FLRW}$  \cite{kkl}. Much of the LQC literature including \cite{aan3} works with such states.}
(Thus, the coefficients of $\t{g}_{ab}$ depend on $\hbar$.) As with the propagation of photons in a medium, this is possible because the cosmological perturbations do not see all the details of quantum fluctuations in $\Psi_{0}$ and the information they do see is captured in $\t{g}_{ab}$. 

One would expect a similar result to hold during the adiabatic phase of black hole evaporation now under consideration since the change in the background curvature occurs very slowly.
 During this phase, the quantum state  $\Psi_{0}({\rm geo})$ of geometry can be taken to be sharply peaked around a smooth metric, and the propagation of the partner modes of the scalar field $\h\Phi$ on this quantum geometry should be well approximated by their propagation on a smooth \emph{dressed effective} metric $\t{g}_{ab}$, constructed from $\Psi_{0}({\rm geo})$. 
 As in the two examples discussed above, the partner modes will not be sensitive to all the detailed quantum fluctuations in $\Psi_{0}({\rm geo})$ and the information they do see will be captured in the dressed effective metric $\t{g}_{ab}$ which is smooth but inherits $\hbar$-dependent terms from $\Psi_{0}({\rm geo})$. There is a  conceptually streamlined  procedure to find this metric. Note that  this $\t{g}_{ab}$  would likely be more refined than the effective metric in the LQG description of Kruskal space-time discussed in Section \ref{s2.3.1}, as it will likely incorporate some of the quantum fluctuations in geometry and the scalar field $\hat{\Phi}$  that are not registered in the effective metric. \vskip0.1cm 

Thus, the study of dynamics in the full trapped region can be divided into three phases: \vskip0.1cm
 (i) A \emph{semi-classical phase} which is expected to provide an excellent approximation to full quantum dynamics in the part of the trapped region that lies to past of the surface $\Ssc$. This is the part of the trapped region to the past of shaded (pink) portion in Fig. 2(b).  During this phase dynamics would be well approximated by quantum field theory of the field $\h\Phi$ on the Vaidya background. During this phase, space-time curvature would be less than $10^{-6}\lp^{2}$.  \vskip0.05cm
(ii) An \emph{adiabatic quantum gravity phase} in which the space-time curvature is larger and enters the Planck regime but the evaporation process is adiabatic. This phase, depicted by the shaded pink region of Fig. 2(b), could be well-described by a dressed effective metric $\t{g}_{ab}$. The task is to calculate this $\t{g}_{ab}$ and use it to describe the propagation of the field $\h\Phi$. This region is  expected to constitute a neighborhood of the transition surface, $\tau$, to the past of which we have a trapped region and to the future of which, an anti-trapped region, both defined using $\t{g}_{ab}$. The metric $\t{g}_{ab}$ will incorporate two distinct sets of quantum effects. The first set is from LQG proper --such as negative effective energy density which is very large in the Planck domain and reduces the mass of the infalling scalar field enormously even at the left end, $v=0$, of the shaded (pink) region, even though the influx of the Hawking partner modes is negligible there. As $v$ increases, we slide to the right in this region, and then the negative energy density due to quantum geometry effects decreases but is compensated by the  negative energy carried by more and more partner modes. The second set of quantum effects is precisely that from dynamics of the scalar field. It includes both, the evolution of the incoming shell and the infalling partner modes. The actual calculation will incorporate both these types of effects in one go. But conceptually it is useful to separate the two contributions, since the first has been absent in most other approaches as they  focus just on the partner modes and also ignore the quantum geometry effects. Finally, the Kruskal results depicted in Fig. 4(a) suggest that the first set of effects will decay rapidly as we move away from Planck curvature  into the semi-classical region. Therefore in the semi-classical region, $\t{g}_{ab}$ will be well approximated by the semi-classical Vaidya metric used there. \vskip0.05cm
(iii) A \emph{full quantum gravity phase} that is localized in the region depicted by a dark (red) blob at the right end of the shaded region in Fig. 2(b). This is the region in which not only is the curvature of Planck scale, but it is varying rapidly because of the evaporation process; dynamical nature of the back reaction is now significant. Therefore, description in terms of $\t{g}_{ab}$ would now be inadequate.  In this phase one would have to use the LQG quantum state for the combined system, adapted to the spherical collapse under consideration. In terms of space-time geometry, the time-like piece of \TDH continues to shrink in the semi-classical region and terminates in the full quantum region as depicted in Fig. 2(b).
\vskip0.1cm

Next, consider the untrapped region to the future of the transition surface $\tau$, defined by the dressed effective metric $\t{g}_{ab}$. Experience with effective equations in the Kruskal extension suggests that the untrapped region will end, its future boundary being an anti-trapping dynamical horizon, \emph{AT-DH}  (see Fig. 2 (b)).  As time evolves, the infrared modes in the trapped region will cross $\tau$ and eventually escape to infinity. Therefore, there would be an outward flux across  \emph{AT-DH}.  Considerations of Section \ref{s2.1} then lead us to the conclusion that \ATDH will be space-like, with its area shrinking in the direction of the projection of $n^{a}$ on \emph{AT-DH}, i.e., in the inward direction.  Finally,  recall that the radius $r$ of 2-spheres achieves its minimum on the transition surface $\tau$, since it increases as we move away from $\tau$ in both future and past directions. Using this fact, one can argue that  the anti-trapping horizon \ATDH will emerge from the fully quantum region and then move left, shrinking in its area and eventually plunging into $\tau$ at its left end, achieving minimum possible area. Note incidentally that the space-time geometry is far from being symmetric around the transition surface. In particular, while the mass $M_{\TDH}$ of the time-like piece of \TDH  decreases from $M_{\odot}$ to some $\sim 100 \mp$, the mass $M_{\ATDH}$ decreases from $\sim 100\mp$ to, say, $\sim 1 \mp$.  However, this asymmetry is present --albeit to a much smaller extent-- already in the LQG extension of Kruskal space-time \cite{aos} discussed in Section \ref{s2.3.1}.

Finally, the region to the future of \ATDH would also be well approximated by a Vaidya metric, but now the outgoing one, expressed in terms of the retarded time coordinate $u$ in place of the advanced time coordinate $v$ of Eq. (\ref{metric}). It will describe the propagation of the infrared modes that will emerge from the \ATDH and arrive at $\scrip$ at very late times. General considerations by Ori \cite{ori4}, and Bianchi et al \cite{ebtdlms,bcdhr}  suggest that the time scales associated with this process will be long, $\mathcal{O}(M^{4}_{\rm ADM})$. 

\subsection{Summary: The global picture}
\label{s2.4}

Let us summarize the discussion of the last three sub-sections. One begins with a coherent state of a quantum scalar field $\h{\Phi}$, peaked around an infalling  classical pulse on $\scrim$\!, undergoing a `prompt collapse'. In the \emph{initial stages} of the collapse, space-time is well approximated by classical GR where  space-time is flat to the past of the support of the pulse, and Schwarzschild to the future, as depicted in Fig. 3(a).  When the radius of the (thick) shell has become sufficiently small, a trapping dynamical horizon \TDH forms. In classical GR of Fig. 3(a), \TDH has only a space-like component. In the quantum theory, it has two components as shown in Fig 2(a). The inner component is space-like, (essentially) the same as in the classical theory, while the outer component is new; it replaces the event horizon  \emph{EH} of Fig. 3(a). However it is time-like, rather than null, and decreases in area due to the negative energy flux of the Hawking `infalling partner modes' as in Fig. 2(a).

The two branches of the \TDH serve as the past boundary of a trapped region. Fig. 2(a) shows the part of the evolution during which semi-classical approximation of Eq. (\ref{semi-class}) holds.  As discussed in Section \ref{s2.2}, during this very long phase of evolution, the mass $M_{\TDH}$ of the time-like part of the \TDH shrinks from the ADM mass, $M_{\rm ADM}$, to  $\sim 9\times 10^{2} \mp$. Pairs of modes are created, one going to infinity and the other falling into the trapped region. The outgoing modes carry away almost all the initial mass to $\scrip$ and the radiation is approximately thermal there. Thus the Bondi mass at $u=u_{LR}$ when the last ray from the semi-classical region reaches $\scrip$ is only of the order of $\sim 9\times 10^{2} \mp$. However, the outgoing modes are entangled with the infalling ones. While the total quantum state $\Psi$ of the system is a functional of both the geometry and the scalar field,  geometry is well-approximated  by a solution to Eq. (\ref{semi-class}) in this semi-classical phase;  it  can therefore be taken to be a smooth metric. Thus, one can focus just on correlations between the (observables associated with the) scalar field in the untrapped region and the trapped region on a Cauchy surface, such as $\Sigma$ of Fig. 2(a). Note however, that the trapped region includes not only the partner modes \emph{but also the infalling coherent state}. The mass associated with the outgoing modes is $(M_{\rm ADM} - M_{\rm Bondi}(u))$, which is huge if $u$ is taken to be sufficiently late so that $M_{\rm Bondi}(u) $ is still macroscopic but much smaller than $M_{\rm ADM}$. In section \ref{s2.2}, as a concrete example, we took  $M_{\rm ADM}$ to be the solar mass $M_{\odot}$ and  $M_{\rm Bondi}(u)$ to be the lunar mass $M_{L}$.  Then mass associated with the part of the state in the trapped region is only $\sim 10^{-7}$ times the mass associated with the outgoing modes. Therefore there is an apparent tension: How can the trapped region house the same number of partner modes as those that went out to $\scrip$ with a  tenth of a million in energy budget? Alternately, since the radius of (the time-like branch of)  \TDH has shrunk to $10^{-2}$cm (corresponding to the $M_{\TDH} = M_{L}$), how can so many modes reside in an apparently tiny region? As we saw in section \ref{s2.2} the answer lies in the fact that although the radius at the \TDH is tiny, restriction of the Cauchy surface $\Sigma$ to the trapped region has an astronomically long neck  --$\sim10^{64}$ light years if $\Sigma$ is taken to be $\Kr = \rm{const}$ surface in the trapped region, and $\sim 10^{62}$ light years if it is a $r=\rm{const}$ surface. Therefore the scalar field modes would be enormously stretched and become infrared, each carrying such a tiny energy that the small energy  budget can be easily accommodated.  I believe that these considerations can be significantly sharpened using quantum field theory on the Vaidya background in the trapped region.%
\footnote{This and other open issues in this program (described below) are being analyzed in collaboration with Tommaso De Lorenzo. Several steps in narrowing down the a priori freedom in the program grew out of  our discussions.}

To the future of the semi-classical region, curvature can exceed $10^{-6} \lp^{-2}$, whence we need quantum gravity. To describe this phase of dynamics, we are guided by two considerations:  the quantum corrected, effective LQG equations describing  Schwarzschild-interior (discussed in section \ref{s2.3.1}), and suitable approximations during the phase in which quantum geometry changes adiabatically (mentioned in section \ref{s2.3.2}). The effective LQG equations imply that the Schwarzschild singularity is resolved and replaced by a transition surface $\tau$, to  the past of which we have a trapped region and to the future, an untrapped region. Curvature is bounded above and reaches its upper bounds on $\tau$.  Experience with other systems --such as the propagation of the quantum electromagnetic field in a medium, or, of cosmological perturbations on the quantum FLRw geometry-- suggests that so long as the evaporation process is adiabatic, i.e., the mass loss is not too fast, one should be able to introduce a smooth metric $\t{g}_{ab}$ that captures all the information in the quantum state of geometry that is seen by the evolving scalar field $\h\Phi$. We are primarily interested in this evolution in order to determine whether the correlations between the early time outgoing and infalling modes are ultimately restored.

Thus, the quantum evolution to the future of the semi-classical region can be divided into two parts: (i)  an \emph{adiabatic quantum gravity phase} in which we continue to have a slowly varying smooth metric $\t{g}_{ab}$ that dictates dynamics of $\h\Phi$; and (ii) a \emph{full quantum gravity phase} in which we have to use the full quantum state $\Psi$ that is now a functional both of geometry and the scalar field. The first of these two phases lasts very long as indicated by the shaded (pink) region in Fig. 2(b), while the second phase is short, as indicated by a dark (red) blob on Fig. 2(b).  There is a systematic procedure to analyze the adiabatic phase in detail and this task can be undertaken soon. Analysis of the full quantum gravity phase is the hardest of open issues since it will involve full LQG as applied to this model. However, it seems reasonable to assume \emph{initially}  that major effect of this full quantum region on future evolution would be in the region bounded by the two null rays $u=u_{1}$ and $u=u_{2}$ of Fig. 2(b). Under this hypothesis one can solve for dynamics in the rest of space-time and see if the expectations about purification at late times are borne out.

Specifically, the proposal is that there will be a transition surface, $\tau$, now in the geometry defined by $\t{g}_{ab}$, separating a trapped and an anti-trapped region. The anti-trapped region will be bounded in the future by an anti-trapping dynamical horizon \emph{AT-DH} (see Fig. 2(b)).  The infrared modes with a very small total energy will leak out across \ATDH and propagate to $\scrip$. The geometry to the future of this \ATDH will be well approximated by the Vaidya metric in the retarded Eddington-Finkelstein  coordinates $u,r,\theta,\varphi$. All along, these modes will  be entangled with those that reach $\scrip$ before $u=u_{1}$ in Fig. 2(b). Therefore purification will occur at late times.  Detailed calculations should reveal if the correlations will be (in essence) fully restored, or if the influence of quantum region between $u=u_{1}$ and $u=u_{2}$ plays a major role in this purification process. There is a  suggestions in LQG \cite{perez} that the ultraviolet structure of quantum geometry may trap some correlations.  Above discussion suggests that the infrared modes will carry significant or almost all correlations but does not rule out the possibility of trapping some correlations in the quantum region.

Analysis of the fully quantum region is the most difficult of open problems. One promising approach is to generalize the analysis of \cite{ori5}  in which the quantum evolution across the singularity of a 2-dimensional black hole was studied. But the required generalization would be quite non-trivial. One may therefore ask: Is the `full quantum gravity phase' essential? It is true that the arguments that enable one to describe quantum geometry using $\t{g}_{ab}$ break down in this region. But could we not perhaps introduce other arguments and describe the entire quantum gravity phase in terms of some smooth metric with coefficients that depend on $\hbar$? Is there an obstruction for such a metric to exist? Detailed considerations involving the transition surface, the behavior of the area radius $r$, and the nature of the trapping and anti-trapping horizons --\TDH and \ATDH-- of Fig. 2(b) imply that such a metric cannot exist in the entire quantum gravity region within the paradigm we have sketched. A `fully quantum gravity phase' is  necessary also from the perspective of radiation at $\scrip$. During the semi-classical phase, as the \TDH shrinks, the temperature associated with the radiation at $\scrip$ grows and more and more \emph{short} wavelength modes are received. On the other had, the radiation that emerges from the anti-trapped region has \emph{extremely long wavelengths} at $\scrip$.  A `fully quantum gravity phase' may play an essential role in understanding this dramatic transition. Thus, there are three seemingly independent considerations that lead one to the conclusion that there must be a genuine quantum gravity region denoted by a (red) blob at the right end of $\tau$: failure of the adiabatic behavior towards the end of the evaporation process, geometric considerations near the transition surface, and dramatic change in nature quantum radiation at $\scrip$.%
\footnote{The second of these reasons --space-time geometry considerations-- seem to apply also to the left end of $\tau$, suggesting that one may have to use quantum geometry also in a blob near the left end.  However, since the first and the third considerations do not apply there, in the main text I focused on the right end.}

\section{Discussion}
\label{s3}

As in most of the investigations of  the issue of information loss in LQG, the conceptual underpinning of the program summarized here can be traced back  to a general  paradigm \cite{aamb}, inspired by the natural resolution of singularities of GR in LQG. This paradigm is based on two observations:  (i) one needs to shift emphasis from event horizons to dynamical horizons; and, (ii)  if the singularity is resolved, quantum space-time would not end where the classical singularity occurred, whence there is `more room' for restoration of correlations. While the paradigm opened an avenue for this restoration, it did not provide a detailed mechanism. Developments in subsequent years have helped  construct the mechanism. First, we now have  detailed analyses of the space-time geometry in the trapped region of a black hole resulting from the gravitational collapse of a null, thick shell, incident from $\scrim$ \cite{mccr, ao,cdl}. Second, we have a much better understanding of the causal structure of the LQG extension of the Schwarzschild-interior \cite{aoslett, aos}. The program outlined in section \ref{s2.4}  incorporates these developments  and outlines concrete steps that are needed to address the issue of whether correlations are restored at $\scrip$. 

There is vast literature on the issue of information loss. So I will focus only on those investigations that are directly relevant to the issues discussed here and compare and contrast them with the ideas discussed in Section \ref{s2}. First, as mentioned in Section \ref{s1}, there are approaches in which one does not expect singularity resolution. For example, using the AdS/CFT conjecture it has ben argued that the Schwarzschild-like singularities will not be resolved in string theory \cite{eh}. If this is indeed the case, then one would be led to a space-time diagram like Fig. 1(b). As already remarked,  I share the general consensus in the GR circles that in this case the evolution from $\scrim$ to $\scrip$ will not be unitary and correlations will be lost. Generally,  the string theory literature considers black holes in presence of a negative cosmological constant as a simplified mathematical context  to probe what would happen in the asymptotically flat case of direct physical interest. However, the arguments tend to make a strong use of the asymptotically anti-de Sitter boundary conditions and it is far from being obvious how they would then extend to the case of direct physical interest. Indeed, it is not often that one finds space-time diagrams analogous to Fig. 2(b) that explicitly provide the hoped-for purification mechanism in the asymptotically flat context. 

However,  discussions motivated by the AdS/CFT considerations do have a key point in common with the LQG based ideas: It is expected that correlations will be restored and the S-matrix would be unitary.  There is a possibility that the singularity is resolved but the result is a piece of space-time with a \emph{new} asymptotic region of its own. In such a case, although the information would not be lost globally, it would be lost from the asymptotic region of the part of the space-time in which the collapse occurred \cite{uw}.  The program discussed here does not favor this possibility but does not rule it out either. This is because of the `fully quantum region', depicted by a dark (red) blob at the right end of the transition surface $\tau$ of Fig. 2(b). However, as indicated in section \ref{s2.4}, the feasible detailed calculations may suffice to reveal  whether `most' --if not all-- of the correlations are restored when the infrared modes traverse the anti-trapped region and emerge at $\scrip$. If they are, then the  `fully quantum region' would not be that relevant for the issue and S-matrix would be unitary. There will be no gross information loss  to another asymptotic region, nor to ultraviolet microscopic degrees of freedom as suggested in \cite{perez}. 

There is a long series of works (see, e.g., \cite{hayward,frolov,bardeen,ebms,crfv,ebtdlms,hhcr,bcdhr,mdcr}) that posit  a space-time structure for the entire process and work out its consequences. Because there are strong consistency conditions on space-time geometry, these analyses have provided valuable insights into what can and cannot happen. 
For example, in some models we know that, if the purification does occur, then certain tame physical requirements imply that the time scale for purification would be $M_{\rm ADM}^{4}$ \cite{ebtdlms,bcdhr}. By and large the focus in these works is on solving \emph{classical} Einstein's equations (with suitable stress-energy tensors) in various patches, and joining them consistently.  Nonetheless the space-time diagram proposed recently in one of these works \cite{mdcr} is similar to Fig. 2(b), although in that work the singularity persists. In most of these works,
energy flow is  monitored, and often one also ensures that violations of energy conditions are all within the bounds known from quantum field theory. However,  in these approaches the issue of \emph{quantum} correlations is rarely considered and so the issue of `purification' that is central to many investigations --including the program presented here-- is not addressed.

Finally, there is much discussion on the issue of young versus old black holes, and long lived remnants. In the program presented here, there is indeed a key difference between a young and an old black hole. Let us compare two lunar mass black holes -- a young one that has just formed from a gravitational collapse and an old one what started out as a solar mass black hole and then evaporated down to the lunar mass, as discussed in Section \ref{s2.2}. While their dynamical horizons will have the same radius $\sim 10^{-2}$cm, and mass $M_{\TDH} = M_{L}$, their  external environment  as well as internal structure will be \emph{very} different. In the second case, the evaporation process would have gone on for some $10^{64}$ years. Therefore, there will be a very large number of outgoing Hawking modes in the exterior region, and an equal number of ingoing modes in the trapped region, the two being entangled. In particular, the area of dynamical horizons will not be a measure of the entropy of what is in the interior (or exterior). However, in both cases the area can be the measure of the surface degrees of freedom of the horizon, i.e., degrees of freedom that can communicate both the outside and inside regions.  Now, because old black holes has so many modes in the interior, it has been argued that it should be possible to produce ones with low mass copiously in particle accelerators and car collisions. But  such processes would be constrained not only by  the energy and momentum conservation laws normally considered in scattering processes in flat space, but also by space-time considerations \cite{ori1}. We saw in section \ref{s2.2} that old black holes have astronomically long interiors. Their creation in car collisions would involve truly baffling changes in the structure of space-time geometry, occurring almost instantaneously.  Analysis of such processes lies well beyond tools that have been used to analyze the possibility of their appearance in accelerators and car collisions.

I would like to conclude by pointing out a key limitation of the program discussed here, and indeed all programs aimed at understanding the information loss issue that I am aware of.  All these discussions are focus on black holes that have a space-like singularity in the classical theory. However, stability analysis of Kerr space-time suggests that the inner horizon would be unstable to perturbations and so a generic black hole singularity would be null. So far, the fate of null singularities has not been analyzed in LQG. The avenue to recovery of correlations pursued here will not be viable for generic black holes unless null singularities are also resolved. 

\section*{Acknowledgments}
I am grateful to Amos Ori,  Tommaso De Lorenzo,  Eugenio Bianchi and Madhavan Varadarajan for discussions that have played a seminal role in shaping my understanding of black hole evaporation over the years.  I would like to thank  Franz Pretorius and Fethi Ramazanoglu for discussions on CGHS black holes, and Javier Olmedo and Param Singh for discussions on the quantum extension of the Kruskal space-time. In addition, I am indebted to Tommaso De Lorenzo for preparing figures.  This work was supported by the NSF grant PHY-1806356, the grant UN2017-9945 from the Urania Stott Fund of the Pittsburgh Foundation, and the Eberly Chair funds of Penn State.


\begin{thebibliography}{99}

\bibitem{swh} S.~W.~Hawking, Particle creation by black holes, Commun. Math. Phys. \textbf{43}, 199-220 (1975).

\bibitem{swh2} S.~W.~Hawking,  Breakdown of predictability in gravitational collapse, Phys. Rev.  D\textbf{14} 2460- 2473  (1976).

\bibitem{hps} S.~W.~Hawking, M. J. Perry and A. Strominger, Soft hair on black holes,  Phys. Rev. Lett. \textbf{116}, 231301 (2016).

\bibitem{uw} W.~G.~Unruh and R.~M.~Wald, Information loss, Rep. Prog. Phys. \textbf{80},   092002 (2017).

\bibitem{marolf} D.~Marolf,  The black hole information problem: past, present, and future, Rep. Prog. Phys. \textbf{80},   092001 (2017).

\bibitem{apms} A.~ Almheiri, D.~Marolf, J.~ Polchinski, J.~ Sully, Black Holes: Complementarity or Firewalls? \texttt{arXiv:1304.6483}.

\bibitem{giddings} S.~Giddings,   Modulated Hawking radiation and a nonviolent channel for information release \texttt{arXiv:1401.5804}.

\bibitem{ak1} A.~Ashtekar and B.~Krishnan, Dynamical Horizons: Energy, Angular Momentum, Fluxes and Balance Laws,  Phys. Rev. Lett. \textbf{89} 261101 (2002). 

\bibitem{ak2} A.~Ashtekar and B.~Krishnan, Dynamical Horizons and Their Properties, Physical
Review D \textbf{68}, 104030 (2003).

\bibitem{bbgvdb} I.~Booth, L.~Brits, J.~A.~Gonzalez and C.~ Van Den Broeck, Marginally trapped tubes and dynamical horizons, Class. Quant. Grav. \textbf{23}, 413-440 (2006).

\bibitem{akrev} A.~Ashtekar and B.~Krishnan, Isolated and Dynamical Horizons and Their Properties, Livi. Reviews (Relativity) \textbf{7}, No 10 (2004).

\bibitem{boothrev} I.~Booth, Black hole boundaries,  Can. J. Phys. \textbf{83} 1073-1099 (2005). 

\bibitem{ao} A.~Ashtekar and A.~Ori (unpublished calculations, 2014), reported in part 2 of \cite{aa-ilqg}.

\bibitem{aa-ilqg} A.~Ashtekar, The issue of information loss: The current status, ILQG seminar of 
February 23rd, 2015, http://relativity.phys.lsu.edu/ilqgs/ashtekar022316.pdf

\bibitem{page} D.~Page,Information in black hole radiation,Phys. Rev. Lett. \textbf{71}, 3743Ð3746 (1993).

\bibitem{asrev} A. ~Ashtekar and P. ~Singh,  Loop Quantum Cosmology: A Status Report,  Class. Quant. Grav.  \textbf{28},  213001 (2011). 

\bibitem{iapsrev} I.~Agullo and P.~Singh, Loop quantum cosmology: A brief review, In \emph{Loop Quantum Gravity: The first 30 years}, edited by A.~Ashtekar and J.~Pullin,  (World Scientific, Singapore, 2017). 

\bibitem{aoslett}  A.~Ashtekar, J.~Olmedo and P.~Singh, Quantum transfiguration of Kruskal black holes, Phys. Rev. Lett. \textbf{121}, 241301 (2018).
 
\bibitem{aos} A.~Ashtekar, J.~Olmedo and P.~Singh, Quantum extension of the Kruskal space-time, Phys. Rev. D\textbf{98}, 126003 (2018).

\bibitem{ori4}A.~Ori, personal communication.

\bibitem{aamb} A.~Ashtekar and M.~Bojowald, Black hole evaporation: A paradigm, Class. Quant. Grav. \textbf{22}, 3349-3362 (2005).

\bibitem{ori1} A.~Ori, Firewall or smooth horizon? Gen. Relativ. Gravit. \textbf{48}, 9 (2016). 

\bibitem{eh} N.~Englehardt and G.~ T.~ Horowitz,  New insights into quantum gravity from gauge/gravity duality,  Int. J. Mod. Phys.   D\textbf{25},  1643002 (2016).

\bibitem{hayward} S.~A.~Hayward, Formation and evaporation of regular black holes, Phys.Rev.Lett. \textbf{96}, 031103 (2006).

\bibitem{frolov} V.~P.~Frolov, Information loss problem and a Õblack holeÔ model with a closed apparent horizon, JHEP \textbf{1405}, 049, (2014).

\bibitem{bardeen}  J.~M.~Bardeen, Black hole evaporation without an event horizon, \texttt{arXiv:1406.4098}.

\bibitem{crfv} C.~Rovelli and F.~Vidotto,Planck stars,  \texttt{arXiv:1401.6562.}
 
\bibitem{ebms} E.~Bianchi and M.~Smerlak, Last gasp of a black hole: unitary evaporation implies non-monotonic mass loss, Gen. Relativ. Gravit. \textbf{46}, 1809 (2014) .

\bibitem{ebtdlms} E.~Bianchi, T.~De Lorenzo and M.~Smerlak, Entanglement entropy production in gravitational collapse: covariant regularization and solvable models, JHEP \textbf{06} 180, (2015).

 \bibitem{hhcr}  H.~M.~Haggard and C.~Rovelli, Black hole fireworks: quantum-gravity effects outside the horizon spark black to white hole tunneling, Phys. Rev. D\textbf{92}, 104020  (2015).

\bibitem{bcdhr} E.~ Bianchi, M.~ Christodoulou, F.~D'Ambrosio, H.~M.~Haggard, and C.~Rovelli, White holes as remnants: A surprising scenario for the end of a black hole,  
 \texttt{arXiv:1802.04264}. 

\bibitem{mdcr}P.~Martin-Dessuad, C.~Rovelli, Evaporating black to white hole, \texttt{arXiv:1905.0251v2}

\bibitem{ori2}A.~Ori,  Approximate solution to the CGHS field equations for two-dimensional evaporating black holes Phys.Rev. D\textbf{82}, 104009 (2010). 

\bibitem{aprlett} A.~Ashtekar, F. Pretorius and F. Ramazanoglu, Surprises in the evaporation of two dimensional black holes, Phys. Rev. Lett. \textbf{106}, 161303, (2011).

\bibitem{apr} A.~Ashtekar, F. Pretorius and F. Ramazanoglu, Evaporation of two dimensional black holes, Phys. Rev.  D\textbf{83}, 044040  (2011).

\bibitem{ori3} A.~Levi and A.~Ori, Two-dimensional semiclassical static black holes: Finite-mass correction to the Hawking temperature and outflux,  Phys. Rev. D\textbf{88}, 024024 (2013). 

\bibitem{rggh} R.~Geroch and G.~T.~Horowitz, Asymptotically Simple Does Not Imply Asymptotically Minkowskian, Phys. Rev. Lett. \textbf{40}, 203-206 (1978); Errata:  Phys. Rev. Lett. 40, 350 (1978); Phys. Rev. Lett. \textbf{40}, 483 (1978). 

\bibitem{atv} A.~Ashtekar, V.~Taveras, and M.~Varadarajan, Phys. Rev. Lett. \textbf{100}, 211302 (2008).

\bibitem{swhjs} S.~W.~Hawking and J.~M.~Stewart, Naked and thunderbolt singularities in black hole evaporation Nucl. Phys. B\textbf{400}, 393-415  (1993).

\bibitem{anderson} P.~R.~Anderson, R.~D. ~Clark, A.~Fabbri, and M.~R.~ R.~ Good
Late time approach to Hawking radiation: terms beyond leading order, 
Phys. Rev. D\textbf{100}, 061703 (2019). 

\bibitem{hayward-pop}  S.~Hayward,  The disinformation problem for black holes (pop version), \texttt{arXiv:gr-qc/0504038}.

\bibitem{hayward-conf} S.~Hayward, The disinformation problem for black holes (conference version), \texttt{arXiv:gr-qc/0504037}.

 \bibitem{cdl} M.~Christodoulou and T.~De Lorenzo, On the volume inside old black holes, Phys. Rev. \textbf{D} \textbf{94}, 104002 (2016). 

 \bibitem{mccr}  M.~Christodoulou and  C.~ Rovelli,   How big is a black hole? Phys. Rev. D\textbf{91}, 064046 (2015). 


\bibitem{ab}  A. ~Ashtekar and M. ~Bojowald,  Quantum Geometry and the Schwarzschild singularity,  Class. Quant. Grav. \textbf{23} 391-411 (2006).

\bibitem{lm}  L. ~Modesto, Loop quantum black hole, Class. Quant. Grav. \textbf{23}  5587-5602 (2006).

\bibitem{bv} C.~ G.~ Boehmer and K.~ Vandersloot, Loop quantum dynamics of Schwarzschild interior,  Phys. Rev. D{\bf 76}, 1004030 (2007).

\bibitem{dc} D.~W.~Chiou, Phenomenological loop quantum geometry of the Schwarzschild black hole,  Phys.\ Rev.\ D {\bf 78}, 064040 (2008). 

\bibitem{cgp} M.~Campiglia, R.~Gambini and J.~Pullin, Loop quantization of a spherically symmetric midi-superspaces: The interior problem, AIP Conf. Proc. \textbf{977}, 52-63 (2008).
%
\bibitem{bkd}  J. Brannlund, S. Kloster, A. DeBenedictis, The Evolution of $\Lambda$ Black Holes in the Mini-Superspace Approximation of Loop Quantum Gravity,  Phys. Rev. D {\bf 79}, 084023 (2009).

\bibitem{cs}  A. ~Corichi and P. ~Singh, Loop quantum dynamics of Schwarzschild interior revisited, Class. Quant. Grav. \textbf{33}, 055006  (2016).  


\bibitem{djs}  N.~Dadhich, A.~Joe and P.~Singh, Emergence of the product of constant curvature spaces in loop quantum cosmology,  Class.\ Quant.\ Grav.\  {\bf 32},  185006 (2015).

\bibitem{oss} J.~ Olmedo,  S. ~Saini and P. ~Singh, From black holes to white holes: a quantum gravitational symmetric bounce, Class.\ Quant.\ Grav.\  {\bf 34},  225011 (2017).

 \bibitem{cctr}   J. Cortez, W. Cuervo, H. A. Morales-Técotl, J. C. Ruelas, On effective loop quantum geometry of Schwarzschild interior,  Phys. Rev. D {\bf 95}, 064041 (2017).


\bibitem{abp} E.~Alesci, S.~Bahrami,and D.~ Pranzetti, Quantum gravity predictions for black hole interior geometry, Phys. Lett. B\textbf{797}, 134908 (2019) .

\bibitem{nbetal} N.~Bodendorfer, F.~M.~Mele and J.~M\"{u}nch, Effective quantum extended spacetime of polymer Schwarzschild black hole, Class. Quantum Grav. \textbf{36}, 195015 (2019).

\bibitem{apslett} A.~Ashtekar, T.~ Pawlowski, and P.~ Singh, A.~Ashtekar, T.~ Pawlowski, P.~ Singh, Phys. Rev. Lett. \textbf{96}, 141301(2006).

\bibitem{apsv} A.~Ashtekar, T.~ Pawlowski, P.~ Singh and  K.~Vandersloot,  Loop quantum cosmology of k=1 FRW models, Phys. Rev. D\textbf{75}, 024035 (2007). 

\bibitem{aan3} I.~Agullo, A.~Ashtekar and W.~Nelson, The pre-inflationary dynamics of loop quantum cosmology: Confronting quantum gravity with observations, Class. Quantum. Grav. \textbf{30},  085014 (2013). 

\bibitem{kkl} W.~Kaminski, M.~Kolanowski and J.~Lewandowski, Dressed metric predictions revisited, \texttt{arXiv:1912.02556} (2019).

\bibitem{ori5} D. Levanony and A.~Ori,  Interior design of a two-dimensional semiclassical black hole: Quantum transition across the singularity,  Phys. Rev. D\textbf{81}, 104036 (2010). 
 
 \bibitem{perez} L.~Amadei and  A.~Perez, Hawking's information puzzle: a solution realized in loop quantum cosmology, \texttt{arXiv:1911.00306}

 
\end{thebibliography}
\end{document}